\documentclass[10pt]{article}
\usepackage[utf8]{inputenc}
\usepackage{amsmath,amssymb,graphicx}
\usepackage{hyperref}
\usepackage[T1]{fontenc}
\usepackage{mathpazo}
\usepackage[font=small,labelfont=bf]{caption}
\usepackage{longtable}
\usepackage{verbatim}
\usepackage{amsfonts}
\usepackage{cite}
\usepackage{float}

\usepackage[table,usenames,dvipsnames]{xcolor}
\definecolor{darkerblue}{rgb}{0.2,0.2,0.5}

\usepackage{afterpage}
\usepackage[utf8]{inputenc}

\newcommand{\unit}{\rm}

\newcommand{\bear}{\begin{array}}
\newcommand{\ear}{\end{array}}

\newcommand{\beq}{\begin{eqnarray}}
\newcommand{\eeq}{\end{eqnarray}}
\newcommand{\beqa}{\begin{eqnarray}}
\newcommand{\eeqa}{\end{eqnarray}}

\def\OMIT#1{{}}
\newcommand{\lsim}{\mathrel{\rlap{\lower4pt\hbox{\hskip1pt$\sim$}}
    \raise1pt\hbox{$<$}}}         
\newcommand{\gsim}{\mathrel{\rlap{\lower4pt\hbox{\hskip1pt$\sim$}}
    \raise1pt\hbox{$>$}}}         
\def\met{\ensuremath{E_{\mathrm{T}}^{\mathrm{miss}}}} 

\usepackage{tikz}
\usetikzlibrary{trees}
\usetikzlibrary{decorations.pathmorphing}
\usetikzlibrary{decorations.markings}
\usetikzlibrary{shapes.misc}
\tikzset{
    photon/.style={decorate, decoration={snake}, draw=black},
    wino/.style={draw=redwine},    
    electron/.style={postaction={decorate},
        decoration={markings,mark=at position .55 with {\arrow[draw=black]{>}}}},
    scalar/.style={dashed,postaction={decorate},
        decoration={markings,mark=at position .55 with {\arrow[draw=black]{>}}}},
    gluon/.style={decorate, 
        decoration={coil,amplitude=4pt, segment length=5pt}},
   crosspoint/.style={thick, draw=black!60!green, cross out, inner sep=0pt, minimum width=5pt, minimum height=5pt,},
   separation/.style={dashed},
}

\usepackage{titlesec}

\titleformat{\section}
{\color{darkerblue}\normalfont\Large\bfseries}
{\color{darkerblue}\thesection}{1em}{}
\titleformat{\subsection}
{\color{darkerblue}\normalfont\large\bfseries}
{\color{darkerblue}\thesubsection}{1em}{}

\title{\bf \color{darkerblue} Stealth Supersymmetry Simplified}

\author{JiJi Fan$^{a}$, Rebecca Krall$^{b}$, David Pinner$^{c}$, Matthew Reece$^{b}$, and Joshua T. Ruderman$^{d}$\\
{\em $^a$ Department of Physics, Brown University, Providence, RI 02912}\\
{\em $^b$ Department of Physics, Harvard University, Cambridge, MA 02138, USA} \\
{\em $^c$ Princeton Center for Theoretical Physics,}\\
{\em Princeton University, Princeton, NJ 08544, USA} \\
{\em $^d$ Center for Cosmology and Particle Physics,}\\
 	{\em Department of Physics, New York University, New York, NY 10003}}

\begin{document}
\maketitle

\begin{abstract}

In Stealth Supersymmetry, bounds on superpartners from direct searches can be notably weaker than in standard supersymmetric scenarios, due to suppressed missing energy.  We present a set of simplified models of Stealth Supersymmetry that motivate 13 TeV LHC searches.  We focus on simplified models within the Natural Supersymmetry framework, in which the gluino, stop, and Higgsino are assumed to be lighter than other superpartners.  Our simplified models exhibit novel decay patterns that differ significantly from topologies of the Minimal Supersymmetric Standard Model, with and without $R$-parity.  We determine limits on stops and gluinos from searches at the 8 TeV LHC\@.  Existing searches constitute a powerful probe of Stealth Supersymmetry gluinos with certain topologies.   However, we identify simplified models where the gluino can be considerably lighter than 1 TeV\@.
Stops are significantly less constrained in Stealth Supersymmetry than the MSSM, and we have identified novel stop decay topologies that are completely unconstrained by existing LHC searches.

\end{abstract}

\section{Introduction}
\label{sec:intro}

The discovery of a Standard Model-like Higgs boson at 125 GeV marks a milestone in particle physics. The presence of the Higgs forces us to confront the naturalness puzzle of the weak scale. Supersymmetry (SUSY) is a leading candidate to stabilize the weak scale.  Naturalness requires that at least some of the Standard Model superpartners, such as stops and Higgsinos, have masses near the weak scale~\cite{Barbieri:1987fn,Dimopoulos:1995mi,Pomarol:1995xc,Cohen:1996vb,Papucci:2011wy,Brust:2011tb}.  
The 8~TeV run of the LHC already constrains a significant portion of Natural SUSY parameter space~\cite{Kribs:2013lua,Arvanitaki:2013yja,Krizka:2012ah,Evans:2013jna,Han:2013kga}.
For example, stops have been probed, in some cases, up to masses of about 700~GeV~\cite{Chatrchyan:2013xna, TheCMScollaboration:14011, TheCMScollaboration:13009, TheCMScollaboration:13015, Aad:2014qaa,Aad:2014bva,Aad:2014kra,Aad:2014nra,Aad:2015pfx}.  However, limits on superpartners, such as stops, are highly model-dependent because they are sensitive to their precise decay topology.  Therefore, it is pressing to answer the question: are there signatures of Natural SUSY that haven't been fully explored by  existing LHC data and searches? Explorations along this line can help guide the ongoing LHC Run 2 searches towards a full coverage of Natural SUSY and to obtain a justified evaluation of the naturalness principle experimentally. 

Indeed a variety of SUSY models exist with very different signatures that might hide from the standard searches~\cite{Barbier:2004ez, Burdman:2006tz,Fan:2011yu,Fan:2012jf,Graham:2012th,Alves:2013wra,Graham:2014vya,Ellwanger:2014hia,Heidenreich:2014jpa,Nakai:2015swg}. In this paper, we focus on the ``Stealth SUSY" class of models proposed by a subset of the authors~\cite{Fan:2011yu,Fan:2012jf,Nakai:2015swg}. The simplest realization of Stealth SUSY is achieved within the Minimal Supersymmetric Standard Model (MSSM): the corner of MSSM parameter space where the stop mass is nearly degenerate with the top mass~\cite{Kats:2011it,Fan:2011yu} and the stop decays to the top and a light bino or gravitino ($m_{\tilde N_1} \ll m_{\tilde t} \approx m_t$). This leads to stop pair production events that look very similar to Standard Model $t{\overline t}$ events. They may be distinguished through spin correlations~\cite{Han:2012fw,Han:2013lna} or precision measurements of the top cross section~\cite{Czakon:2014fka}.  Both of these techniques have now been pursued in data, with no signs of new physics so far~\cite{Aad:2014kva,Aad:2014mfk}.\footnote{Note however that existing spin and cross section limits~\cite{Aad:2014kva,Aad:2014mfk} rely on measurements of the top mass that {\it assume} the SM description of the top.  If top mass measurements are biased by the presence of the stop, existing spin and cross section limits can be greatly weakened~\cite{Czakon:2014fka,Eifert:2014kea}.} As a result, we here do not discuss this ``stealth stop'' and instead focus on beyond-the-MSSM models of Stealth SUSY\@. In these extended models, the MSSM is augmented with a light hidden sector which only feels SUSY breaking weakly and consequently is approximately supersymmetric. The simplest hidden sector is just a SM gauge singlet supermultiplet containing a nearly degenerate fermion and scalar. 

In Stealth SUSY, MSSM superparticles decay through the light hidden sector.  Missing energy is naturally small due to an approximate degeneracy, protected by supersymmetry, between a hidden sector fermion and its scalar superpartner.  The hidden fermion decays to the scalar plus the lightest superparticle ({\it e.g.}~a gravitino or axino), whose momentum is soft due to the degeneracy.
The long cascade decay chains of Stealth SUSY lead to high multiplicity final states with softer visible objects than traditional supersymmetry,  in common with the supersymmetric Hidden Valley scenario\cite{Strassler:2006im,Strassler:2006qa,Nakai:2015swg}. These general features could result in a (considerable) weakening of constraints from the standard searches.

A few dedicated searches for Stealth SUSY have been already performed. One focused on events with photons arising from ${\tilde \chi}^0_1 \to \gamma {\tilde S}$ decays. This is a distinctive signature and the limits are very strong~\cite{CMS:2012un, CMS:2014exa}, disfavoring some of the original Stealth SUSY models. A complementary search focuses on events containing ${\tilde \chi}^\pm_1 \to W^\pm {\tilde S}$ decays~\cite{CMS:2014exa} (but vetoes $b$-jets or additional leptons, making it inefficient for the signals we consider in this paper). Finally, a recent search adds important sensitivity in the case of long-lived particles~\cite{Aad:2015rba}. These are promising contributions toward the effort to extend the LHC's SUSY search strategy beyond minimal scenarios. Our goal in this paper is to make it easy for future searches to target the regions of parameter space still left unconstrained or weakly constrained by these initial attempts.

We use simplified models containing a combination of the key ingredients in natural SUSY: gluinos, stops and Higgsinos. The hidden sector is chosen to be a singlet supermultiplet with a fermionic component $\tilde{S}$ and a scalar component $S$. The simplified stealth models of gluinos have been partially studied in~\cite{Evans:2013jna}, which found that the long gluino decay chains leading to a top/$W$/$Z$ rich final state are still strongly constrained by the Run 1 data. We extend their work and confirm their results. In addition, we study simplified stealth models of stop pair production, and we include Higgsinos in decay chains. 

We now summarize our main results.  We find that gluinos are highly constrained if gluino decays produce top quarks, requiring that the gluino be heavier than about 1~TeV\@.  Alternatively, if the gluino decays to light flavor jets and no missing energy, we find that the gluino can be as light as $\sim 400-600$~GeV\@.  In topologies where stop decays produce multiple $b$-jets, stops can be constrained, depending on the singlino mass, to be as heavy as $\sim 400$~GeV\@.  We have also identified two topologies with no LHC limits on stops,  independent of the stop and singlino  masses.  This implies that  Stealth SUSY stops can be {\it stealthier} than stops in $R$-parity scenarios, where paired dijet searches have recently set limits as strong as $\sim 350-400$~GeV~\cite{Khachatryan:2014lpa}.

Although we focus on direct searches in this paper, we note that indirect constraints via Higgs coupling measurements can be a powerful and complementary probe of natural SUSY\@. The light superpartners in natural SUSY can have a considerable contribution to the Higgs properties and are therefore constrained by Higgs coupling measurements~\cite{Arvanitaki:2011ck,Blum:2012ii,Gupta:2012fy,Craig:2013xia,Farina:2013ssa,Kribs:2013lua,Fan:2014txa,Espinosa:2012in}.
Indirect constraints via Higgs coupling measurements are limited in reach ({\it e.g.}, current Higgs data is only sensitive to scenarios with both stops lighter than about 400 GeV~\cite{Fan:2014txa}), but are complementary to direct searches and are independent of the stop decay topology.
These indirect searches could be an especially powerful probe of stealthy natural scenarios if future $e^+ e^-$ colliders are constructed \cite{Craig:2013xia,Henning:2014gca,Craig:2014una,Fan:2014axa}.   

The paper is organized as follows. In Sec.~\ref{sec:simplified}, we present simplified stealth models with some combinations of gluino, stop, and Higgsinos in the spectrum. We list the decays of the new particles beyond the MSSM. In Sec.~\ref{sec:recasting}, we explain our procedure for making use of existing LHC searches in our analysis. In Sec.~\ref{sec:gluinoconstraint} and Sec.~\ref{sec:stopconstraint}, we present constraints on simplified models of gluino and stop pair production, respectively. We conclude and discuss future prospects in Sec.~\ref{sec:outlook}.

\section{Simplified Natural Stealth Models}
\label{sec:simplified}

Simplified models contain a small set of production and decay channels of new particles, in order to isolate a search channel of interest and to avoid the complexity of a complete model like the MSSM\@. In this context, a number of searches that isolate signatures of naturalness, such as decays of stops to neutralinos or charginos, have been considered, and a large amount of theoretical effort has gone into proposing strategies for finding such stops (or sbottoms)~\cite{Meade:2006dw,Kitano:2006gv,Perelstein:2007nx,Perelstein:2008zt,Han:2008gy,Alwall:2008ag,Plehn:2010st,Papucci:2011wy,Brust:2011tb,Kats:2011qh,Essig:2011qg,Plehn:2011tf,Bi:2011ha,Baer:2012uy,Plehn:2012pr,Bai:2012gs,Lee:2012sy,Alves:2012ft,Han:2012fw,Kaplan:2012gd,Kilic:2012kw,Graesser:2012qy,Dutta:2013sta,Buckley:2013lpa,Belanger:2013gha,Low:2013aza,Bai:2013ema,Hagiwara:2013tva,Dutta:2013gga,Buckley:2014fqa,An:2015uwa,Macaluso:2015wja,Kaufman:2015nda,Kobakhidze:2015scd,Belyaev:2015gna}. Many experimental searches have been undertaken based on these natural SUSY simplified models~\cite{Chatrchyan:2013xna, TheCMScollaboration:14011, TheCMScollaboration:13009, TheCMScollaboration:13015, Aad:2014qaa,Aad:2014bva,Aad:2014kra,Aad:2014nra,Aad:2015pfx}.   Despite this significant and important effort, in a scenario like Stealth SUSY very little of the previous work remains applicable. We require a new set of simplified models and a corresponding new set of search strategies. In this section, we present a set of simplified models that can guide a new set of LHC searches. These models capture the essential features of theories that are {\em both} natural and stealthy. 

We consider simplified models containing (a subset of) gluino, stops, and Higgsinos while decoupling all the other superparticles in the MSSM\@. In this paper we only include the right-handed stop in the signals we analyze, although a more complete natural Stealth SUSY model would also consider signals due to the left-handed stop and sbottom. The stealth sector is chosen to be a gauge singlet multiplet with a fermion $\tilde{S}$ and an almost degenerate scalar $S$\@. In addition, the lightest superparticle in the spectrum is a light invisible fermion, e.g.~a gravitino or axino. To specify a simplified model, we need two important ingredients: the decays of the new states ${\tilde S}$ and $S$ and the decay of the Lightest Ordinary Superpartner (LOSP) to the singlino. We  now present the set of decay scenarios we consider.

\subsection{Singlet and Singlino Decays}

We consider two basic scenarios for the decay of the singlet scalar $S$. Either it decays through mixing with the Higgs boson (the $S H_u H_d$ model) or it decays through loops of new vectorlike particles with Standard Model charges (the $SY{\overline Y}$ model)~\cite{Fan:2011yu,Fan:2012jf}. In the former case, its branching ratios will be identical to those the Standard Model Higgs boson would have if its mass were $m_S$ \cite{Dittmaier:2011ti}.  For most of the simplified models, we focus on a benchmark value $m_S = 90$ GeV where the dominant decay is to $b {\overline b}$, but we also include the leading subdominant decays to $\tau^+ \tau^-$, $gg$, and $c {\overline c}$ as well as the subdominant but phenomenologically interesting decay to $\gamma \gamma$. (We ignore other sub-percent level decays to off-shell $W$ and $Z$ bosons, strange quarks, and muons.) In the case of the $S Y{\overline Y}$ model, we include a dominant decay to gluons and a subdominant decay to photons at the branching fraction $4 \times 10^{-3}$ (as computed for a benchmark point in ref.~\cite{Fan:2011yu}). The branching fractions that we assume are listed in Table~\ref{tab:singletdecays}.

\begin{table}[h]
\begin{center}
\setlength{\tabcolsep}{.3em}
\begin{tabular}{|c|c|}
\hline
Singlet Decay & $SH_u H_d$ Branching Fraction\\
\hline
$b {\overline b}$ & 81.4\% \\
$\tau^+ \tau^-$ & 8.3\% \\
$gg$ & 6.1\% \\
$c {\overline c}$ & 4.1\% \\
$\gamma\gamma$ & 0.1\% \\
\hline
\end{tabular}
\quad
\begin{tabular}{|c|c|}
\hline
Singlet Decay & $SY {\overline Y}$ Branching Fraction\\
\hline
$gg$ & 99.6\% \\
$\gamma\gamma$ & 0.4\% \\
\hline
\end{tabular}
\caption{Singlet branching ratios assumed in our simulations. These are intentionally simplified to capture the phenomenological highlights, rather than computed in the complete theory.}  
\label{tab:singletdecays}
\end{center}
\end{table}

Because $S$ is the scalar part of a chiral supermultiplet, it actually contains both real and imaginary (pseudoscalar) parts: $S = \frac{1}{\sqrt{2}} \left(s + i a\right)$. However, for phenomenological purposes, there is little distinction between $s$ and $a$ in the models we consider. In the spirit of simplified models, to capture the relevant collider signals we only discuss one degree of freedom and refer to it as $S$.

We take the singlino to be slightly heavier than the singlet. Thus the singlino will decay 100\% of the time to singlet and the invisible fermion, which is the lightest superparticle in the spectrum. 

\subsection{Decays of the LOSP}

\begin{figure}[H]
\begin{center}
\resizebox{0.6\textwidth}{!}{%
\begin{tikzpicture}[line width=1.5 pt]
\node at (-1.2,0) {$\tilde{g}$};
\node at (-1.2, 2.4) {LOSP};
\node at (2,2.4) {$SH_uH_d$}; 
\draw (0,0)--(2,0);
\node at (1,0.5) {${\tilde g}$};
\draw[gluon] (0,0)--(2,0);
\draw (2,0)--(3.5,1.0);
\node at (3.8,1.2) {$t$};
\draw (2,0)--(3.5,0.0);
\node at (3.8,0.0) {$\bar{t}$};
\draw[red,fill=red] (2,0) circle (1.2ex);
\draw (2,0)--(3.5,-1.0);
\node at (3.7,-1.2) {${\tilde S}$};
\node[crosspoint, rotate=50] at (3.0, -0.65) {};
\draw [separation] (5.0, 3.0) -- (5.0, - 11.0);
\begin{scope}[shift={(5.5,0.0)}]
\node at (2, 2.4) {$SY\overline{Y}$} ;
\draw (0,0)--(2,0);\draw[gluon] (0,0)--(2,0);
\node at (1,0.5) {${\tilde g}$};
\draw[blue,fill=blue] (2,0) circle (1.2ex);
\draw[gluon] (2,0)--(3.5,1.5);
\node at (3.7,1.8) {$g$};
\draw (2,0)--(3.5,-1.0);
\node at (3.7,-1.2) {${\tilde S}$};
\end{scope}
\begin{scope}[shift={(0.0,-3.2)}]
\node at (-1.2,0) {$\tilde{t}$};
\draw [scalar](0,0)--(2,0);
\node at (1,0.5) {${\tilde t}$};
\draw (2,0)--(3.5,1.0);
\node at (3.8,1.2) {$t$};
\draw (2,0)--(3.5,-1.0);
\node at (3.7,-1.2) {${\tilde S}$};
\node[crosspoint, rotate=50] at (3.0, -0.65) {};
\end{scope}
\begin{scope}[shift={(5.5,-3.2)}]
\draw [scalar](0,0)--(2,0);
\node at (1,0.5) {${\tilde t}$};
\draw [gluon] (2,0)--(3.5,1.0);
\node at (3.8,1.2) {$g$};
\draw (2,0)--(3.5,0.0);
\node at (3.8,0.0) {$t$};
\draw[blue,fill=blue] (2,0) circle (1.2ex);
\draw (2,0)--(3.5,-1.0);
\node at (3.7,-1.2) {${\tilde S}$};
\end{scope}
\begin{scope}[shift={(0.0,-6.4)}]
\node at (-1.2,0) {$\tilde{H}^+$};
\draw [electron](0,0)--(2,0);
\node at (1,0.5) {${\tilde H}^+$};
\draw[photon] (2,0)--(3.5,1.0);
\node at (4.0,1.2) {$W^+$};
\draw (2,0)--(3.5,-1.0);
\node at (3.7,-1.2) {${\tilde S}$};
\node[crosspoint, rotate=50] at (3.0, -0.65) {};
\end{scope}
\begin{scope}[shift={(5.5,-6.4)}]
\draw [electron](0,0)--(2,0);
\node at (1,0.5) {${\tilde H}^+$};
\draw [photon] (2,0)--(3.5,1.0);
\node at (4.0,1.2) {$W^+$};
\draw[blue,fill=blue] (2,0) circle (1.2ex);
\draw (2,0)--(3.5,-1.0);
\node at (3.7,-1.2) {${\tilde S}$};
\end{scope}
\begin{scope}[shift={(0.0,-9.6)}]
\node at (-1.2,0) {$\tilde{H}^0$};
\draw [electron](0,0)--(2,0);
\node at (1,0.5) {${\tilde H}^0$};
\draw[photon] (2,0)--(3.5,1.0);
\node at (4.0,1.2) {$Z,h,S$};
\draw (2,0)--(3.5,-1.0);
\node at (3.7,-1.2) {${\tilde S}$};
\node[crosspoint, rotate=50] at (3.0, -0.65) {};
\end{scope}
\begin{scope}[shift={(5.5,-9.6)}]
\draw [electron](0,0)--(2,0);
\node at (1,0.5) {${\tilde H}^0$};
\draw [photon] (2,0)--(3.5,1.0);
\node at (4.0,1.2) {$Z, \gamma$};
\draw[blue,fill=blue] (2,0) circle (1.2ex);
\draw (2,0)--(3.5,-1.0);
\node at (3.7,-1.2) {${\tilde S}$};
\end{scope}
\end{tikzpicture}
}
\end{center}
\caption{All possible LOSP decays we consider in the stealth simplified models. The crosses on the singlino legs in the $SH_uH_d$ models represent a mixing between singlino and Higgsino. The red blob in the gluino LOSP decay in the $SH_uH_d$ model represents an off-shell stop. The blue blobs in all the LOSP decays in the $SY\overline{Y}$ model represent a loop of $Y, \overline{Y}$ messengers. 
} \label{fig:LOSPdecays}
\end{figure}
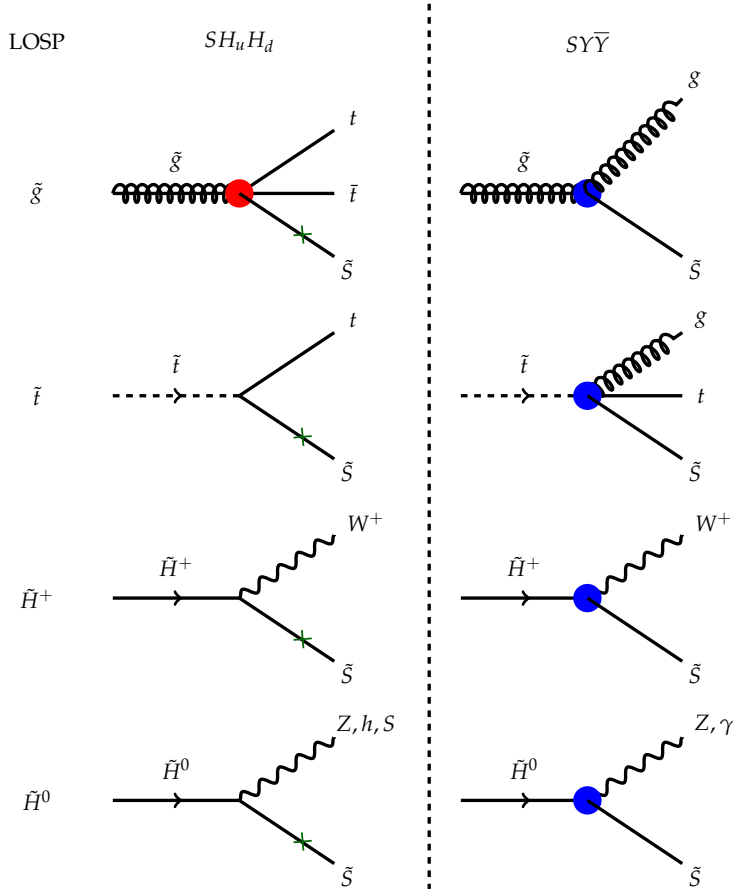

We assume a relatively small coupling between the visible sector and the stealth sector, so that all visible superpartners rapidly cascade down to the LOSP, which then decays to the stealth sector. The decays that we obtain depend on the choice of $SH_u H_d$ or $SY\overline{Y}$ as the dominant interaction between the sectors. Further detailed descriptions of decay widths (including the relevant natural SUSY widths for transitions between SM superpartners) may be found in Appendix \ref{app:branching}. We list the decays of different possibilities of LOSP in either the $SH_u H_d$ or $SY\overline{Y}$ model in Figure~\ref{fig:LOSPdecays}. In the $SH_uH_d$ model, the decays of the LOSP to singlino always go through mixing between the singlino and a Higgsino (represented by the crosses in the figure). In the $SY\overline{Y}$ model, the couplings between the LOSP and the stealth singlet are generated via a loop of $Y$ and $\overline{Y}$ messengers. Thus the decays of LOSP into the stealth sector proceed through the loops represented by the blue blobs in the figure.

We have drawn all of the Higgsino states decaying to the singlino in Figure~\ref{fig:LOSPdecays}, treating the whole multiplet as the LOSP. There are also transitions within the Higgsino multiplet, e.g.~${\tilde H}^+ \to W^{+*} {\tilde H}^0_1$. These transitions are highly phase-space suppressed due to the small splittings within the Higgsino multiplet. On the other hand, the decays to singlino are suppressed by small mixings in the $SH_u H_d$ model and by loop factors in the $SY\overline{Y}$ model. As a result, in different regions of parameter space the dominant decay can be either the usual MSSM transition to the lightest Higgsino (followed by Higgsino decay) or the decay to singlino as in the figure. In our simulations, we always assume that all Higgsinos decay directly to the singlino, but in general both cases are interesting.

\section{Recasting Existing Searches}
\label{sec:recasting}

\begin{figure}[!h]\begin{center}
\includegraphics[width=0.5\textwidth]{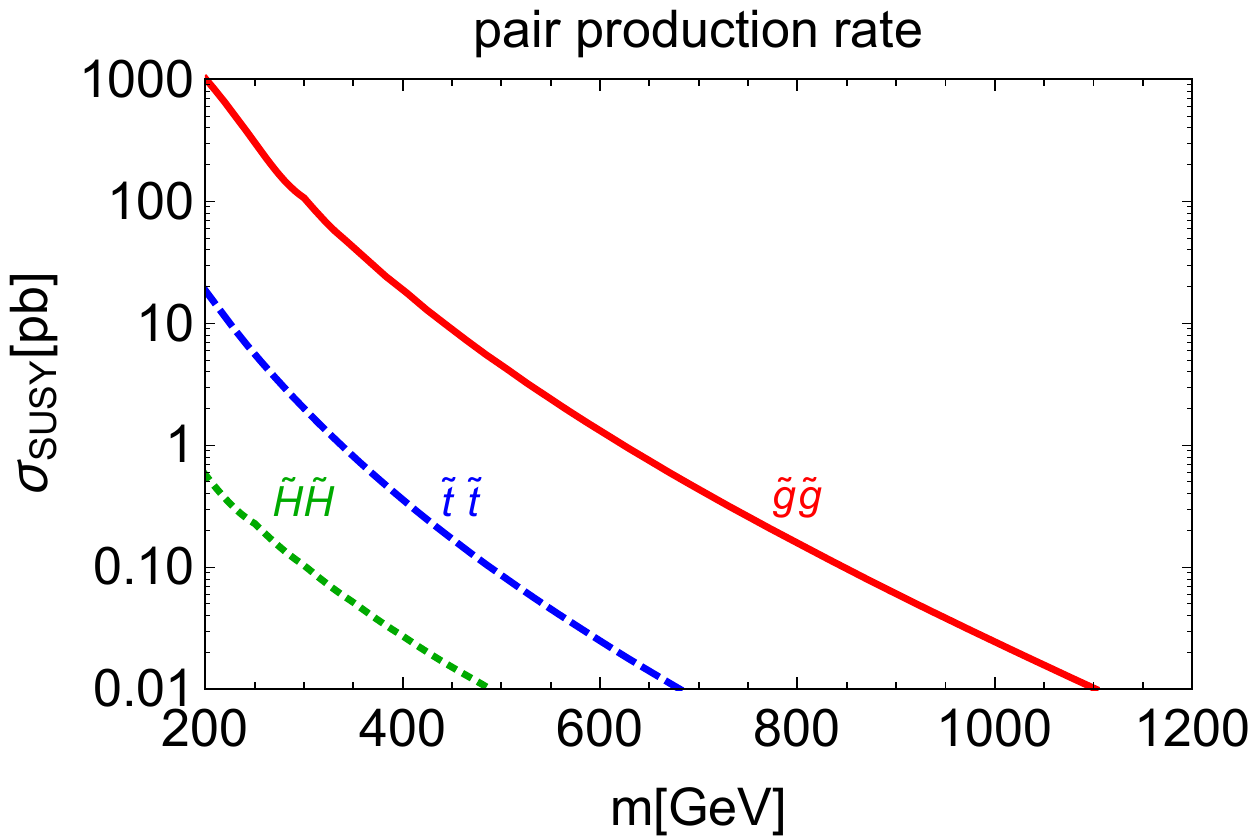}
\end{center}
\caption{Pair production rate of gluinos, stops and Higgsinos at the 8 TeV LHC. The gluino and stop pair production rates are taken from~\cite{Borschensky:2014cia} while the Higgsino pair production rate is taken from~\cite{Fuks:2012qx,Fuks:2013vua}.}
\label{fig:rate}
\end{figure}

We consider pair production of gluinos or right-handed stops for each simplified model. The gluino and stop pair production rates are taken from~\cite{Borschensky:2014cia}. The rates are calculated at the next-to-leading order in the strong coupling constant and include the resummation of soft gluon emission at the next-to-leading logarithmic accuracy. For the reader's convenience, we present the production cross section as a function of the mass of the relevant SUSY particle in Figure~\ref{fig:rate}. Directly searching for electroweak production of Higgsinos is interesting, but given the smaller cross sections we do not consider it in detail in this paper.

We simulated parton level events using MadGraph/MadEvent \cite{Maltoni:2002qb,Alwall:2007st,Alwall:2011uj} and showered the events through Pythia \cite{Sjostrand:2007gs,Sjostrand:2014zea}. In generating the events, we also used LHAPDF for parton distribution functions \cite{Whalley:2005nh}, the SUSY Les Houches accord \cite{Skands:2003cj,Allanach:2008qq}, and HepMC to convert event records\cite{Dobbs:2001ck}. We used the publicly available code CheckMATE (version 1.2.0)~\cite{Drees:2013wra}, which checks the compatibility of new physics models  with a subset of LHC searches. The CheckMATE package relies on DELPHES 3 \cite{deFavereau:2013fsa} and the $M_{T2}$ variable \cite{Lester:1999tx, Barr:2003rg, Cheng:2008hk}. We also selected and coded up some searches that could be sensitive to the simplified stealth models but are not available in CheckMATE. The searches we checked are tabulated in Table~\ref{tab:searches}. We have validated all of our codes and the details of the validation could be found in Appendix~\ref{app:validation}. In the case of the ATLAS multijet study \cite{TheATLAScollaboration:2013xia,Aad:2015lea}, we computed the number of allowed events ourselves using $CL_s$ because we could not reconcile the quoted limits on number of events per channel with the counts that were provided.
Our analysis codes and CheckMATE codes both use the FastJet package\cite{Cacciari:2011ma,Cacciari:2005hq} with the anti-$k_t$ algorithm \cite{Cacciari:2008gp} to cluster jets, and the $CL_s$ prescription to set limits \cite{Read:2002hq}. Ideally it would be useful to have an official recasting from the experiments themselves following the proposal in~\cite{Cranmer:2010hk}. In the following sections, we only present the strongest constraints from the searches we coded up and those included in CheckMATE package. The most constraining searches in the CheckMATE package are tabulated in Table~\ref{tab:checkmate}.

\begin{table}[H]
\begin{center}
\setlength{\tabcolsep}{.3em}
\begin{tabular}{|c|c|c|}
\hline
Search & Data(fb$^{-1}$) &Reference\\
\hline
ATLAS massive particles in multijet channel & 20.3 & \cite{TheATLAScollaboration:2013xia,Aad:2015lea} \\
ATLAS stop search in the $Z + b + \met$ channel&20.3&\cite{Aad:2014mha}\\
ATLAS search with large jet multiplicity and $\met$ &20.3 &~\cite{Aad:2013wta}\\
CMS jet multiplicity distribution in top pair production&19.7 &\cite{Khachatryan:2015oqa} \\
CMS same-sign dileptons and jets&19.5&\cite{Chatrchyan:2013fea} \\
CMS stop searches with $h \to \gamma\gamma$ & 19.7&\cite{Chatrchyan:2013mya} \\
CMS multilepton searches & 19.5 &\cite{Chatrchyan:2014aea}\\
\hline
\end{tabular}
\caption{LHC searches checked in the study in addition to the searches in the public CheckMATE package~\cite{Drees:2013wra}. All of them use 8 TeV data. The details of the validation could be found in Appendix~\ref{app:validation}.}  
\label{tab:searches}
\end{center}
\end{table}

\begin{table}[H]
\begin{center}
\setlength{\tabcolsep}{.3em}
\begin{tabular}{|c|c|c|}
\hline
Search & Data(fb$^{-1}$) &Reference\\
\hline
ATLAS same-sign dilepton and trilepton plus jets & 20.3 &\cite{Aad:2014pda} \\
ATLAS search for missing energy and at least 3 $b$-jets&20.1&\cite{TheATLAScollaboration:2013tha,Aad:2014lra}\\
\hline
\end{tabular}
\caption{The most constraining searches for the stealth models in the CheckMATE package. The first search is validated by us, while the second one is validated already in the CheckMATE package.}  
\label{tab:checkmate}
\end{center}
\end{table}

We have not done a detailed recasting of the existing stealth SUSY search relying on photons or $W$ bosons \cite{CMS:2014exa}. Photon branching ratios are small in the models we consider, and the $W$ search rejected any event with a $b$-tagged jet and required exactly two opposite-sign leptons, making it inefficient for our signals.

\section{Constraints on Gluinos}
\label{sec:gluinoconstraint}

\subsection{The ${\tilde g} \to {\tilde S} \to {\tilde G}$ Decay Chain}

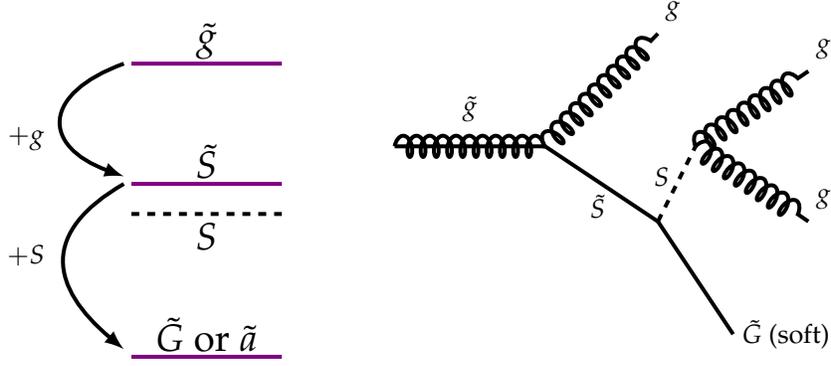
\begin{figure}[H]
\begin{center}
\begin{tikzpicture}[line width=1.5 pt]
\draw[violet] (0,1.6)--(2,1.6);
\node at (1,1.9) {\Large ${\tilde g}$};
\draw[violet] (0,0)--(2,0);
\draw[dashed] (0,-0.4)--(2,-0.4);
\draw[violet] (0,-2.3)--(2,-2.3);
\node at (1,0.3) {\Large ${\tilde S}$};
\node at (1,-0.7) {\Large $S$};
\node at (1,-2.0) {\Large ${\tilde G}~{\rm or}~{\tilde a}$};
\draw[-latex] (-0.1,0) to[out=210,in=140,looseness=1.5] (-0.1,-2.2);
\node at (-1.4,0.6) {$+g$};
\draw[-latex] (-0.1,1.6) to[out=200,in=160,looseness=2] (-0.1,0.1);
\node at (-1.4,-0.95) {$+S$};
\begin{scope}[shift={(3.5,0.5)}]
\draw (0,0)--(2,0);\draw[gluon] (0,0)--(2,0);
\node at (1,0.5) {${\tilde g}$};
\draw[gluon] (2,0)--(3.5,1.5);
\node at (3.7,1.8) {$g$};
\draw (2,0)--(3.5,-1.0);
\node at (2.7,-0.8) {${\tilde S}$};
\draw[dashed] (3.5,-1.0)--(4.0,0);
\node at (3.55,-0.4) {$S$};
\draw[gluon] (4.0,0.0)--(5.5,1.0);
\node at (5.7,1.3) {$g$};
\draw[gluon] (4.0,0.0)--(5.5,-1.0);
\node at (5.7,-0.7) {$g$};
\draw (3.5,-1.0)--(4.5,-2.5);
\node at (5.2,-2.5) {${\tilde G}$ (soft)};
\end{scope}
\end{tikzpicture}
\end{center}
\caption{The ${\tilde g} \to {\tilde S} \to {\tilde G}$ simplified model in the $SY{\overline Y}$ scenario. Effectively, a gluino decays to three gluons (plus a soft particle), so the signal is ${\tilde g}{\tilde g} \to 6~{\rm jet}+{\rm soft}$.
} \label{fig:simplifiedGluinoSinglinoSYY}
\end{figure}

The simplified model for a minimal scenario of a gluino decaying directly through the singlino is depicted in Figure~\ref{fig:simplifiedGluinoSinglinoSYY} for the $SY{\overline Y}$ scenario.   For simplicity, we specialize to the $S Y {\overline Y}$ model in this section.  We do not consider the $S H_u H_d$ model because it generates similar signatures to the ${\tilde g} \to {\tilde t} \to {\tilde S}$ chain of the  $S H_u H_d$ model discussed in the next subsection, as the dominant decay in that case is often ${\tilde g} \to t {\overline t} {\tilde S}$. (If the phase space is insufficient, the loop decay $g {\tilde S}$ may also arise in the $SH_u H_d$ model, rendering it similar to the case discussed here.)

When $m_{\tilde S} = 100$ GeV, we find that gluino masses up to $\sim 400-600$~GeV are excluded at the 95\% confidence level by the ATLAS multijet search~\cite{TheATLAScollaboration:2013xia,Aad:2015lea}. The signal region extends to higher gluino masses in the case of a heavier ${\tilde S}$ mass, in which case there are more hard jets in the signal.  We find that the ATLAS channels with most gluino sensitivity search for events with at least seven non-$b$-tagged jets (with various $p_T$ cuts, all above 100 GeV).

\begin{figure}[H]\begin{center}
 \includegraphics[width=0.45\textwidth]{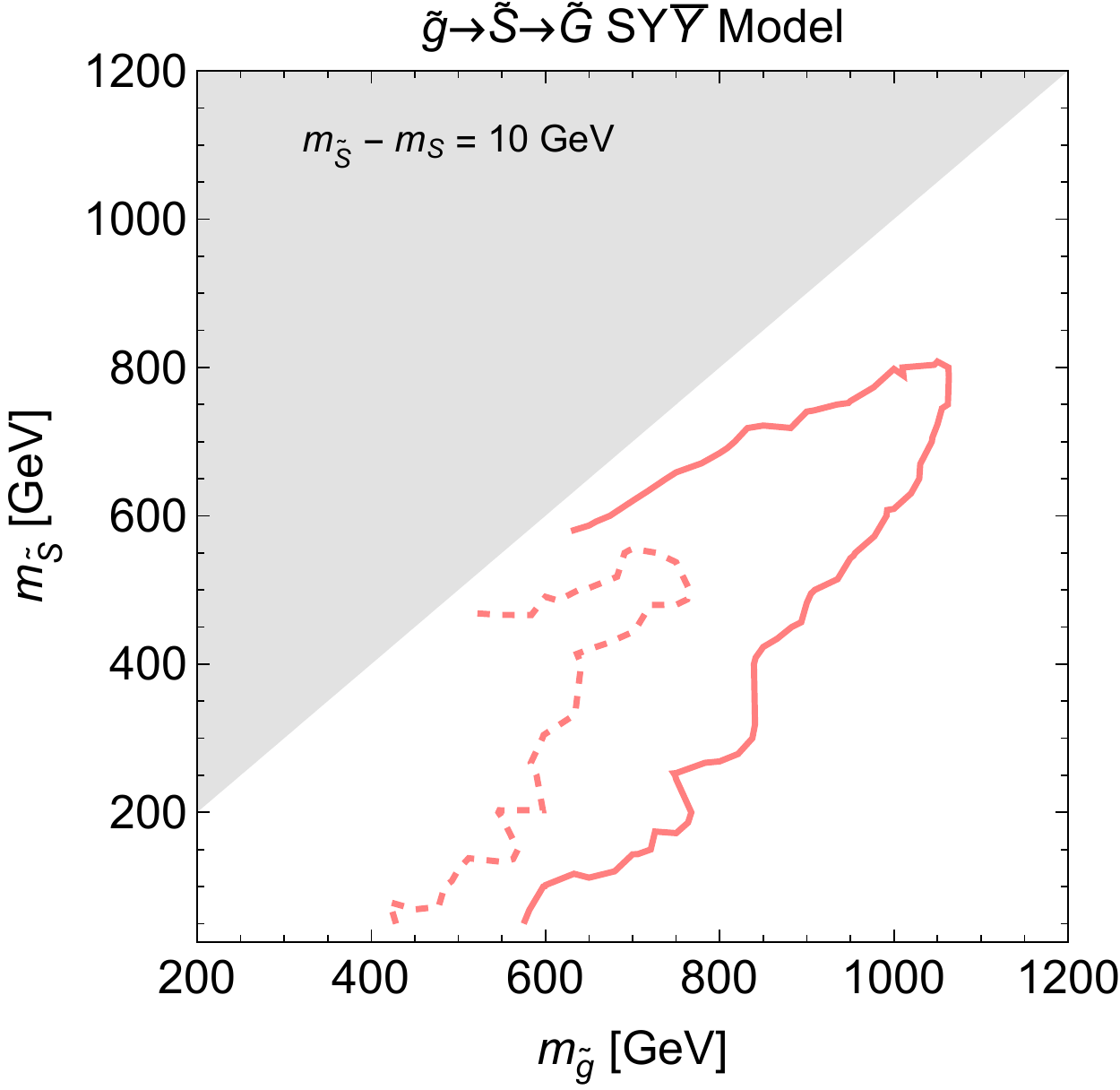} 
\end{center}
\caption{Bound on the ${\tilde g} \to {\tilde S} \to {\tilde G}$ decay chain. The red solid line is the 95\% confidence exclusion from the ATLAS multijet search~\cite{TheATLAScollaboration:2013xia,Aad:2015lea}, and the red dashed line is a conservative constraint in which we weaken the bounds by a factor of 2 to account for possible systematic overestimation of the signal efficiency.}
\label{fig:boundGlStS}
\end{figure}%

\subsection{The ${\tilde g} \to {\tilde t} \to {\tilde S} \to {\tilde G}$ Decay Chain}
\label{subsec:gluinostopsinglino}

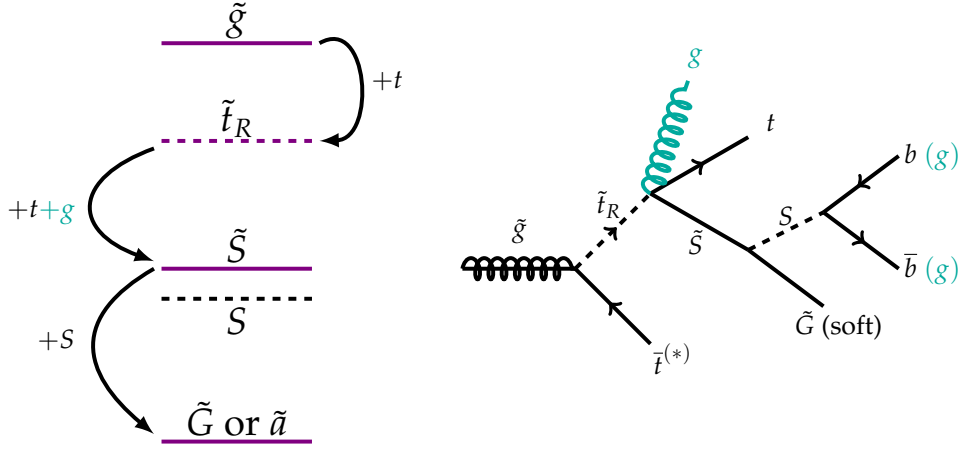
\begin{figure}[!h]
\begin{center}
\begin{tikzpicture}[line width=1.5 pt]
\node at (1,3.3) {\Large ${\tilde g}$};
\draw[violet] (0,3.0)--(2,3.0);
\node at (1,2.0) {\Large ${\tilde t}_R$};
\draw[violet,dashed] (0,1.7)--(2,1.7);
\draw[-latex] (2.1,3.0) to[out=30,in=2,looseness=1.5] (2.1,1.7);
\node at (3.0,2.5) {$+t$};
\draw[violet] (0,0)--(2,0);
\draw[dashed] (0,-0.4)--(2,-0.4);
\draw[violet] (0,-2.3)--(2,-2.3);
\node at (1,0.3) {\Large ${\tilde S}$};
\node at (1,-0.7) {\Large $S$};
\node at (1,-2.0) {\Large ${\tilde G}~{\rm or}~{\tilde a}$};
\draw[-latex] (-0.1,0) to[out=210,in=140,looseness=1.5] (-0.1,-2.2);
\node at (-1.6,0.75) {$+t{\color{Emerald}+g}$};
\draw[-latex] (-0.1,1.6) to[out=200,in=160,looseness=2] (-0.1,0.1);
\node at (-1.4,-0.95) {$+S$};
\begin{scope}[shift={(4.0,0.0)}]
\draw (0,0)--(1.5,0);\draw[gluon] (0,0)--(1.5,0);
\node at (0.76,0.5) {${\tilde g}$};
\draw[scalar] (1.5,0)--(2.5,1.0);
\node at (1.95,0.85) {${\tilde t}_R$};
\draw[electron] (2.5,-1.0)--(1.5,0);
\node at (2.8,-1.2) {${\overline t}^{(*)}$};
\draw[electron] (2.5,1.0)--(3.8,1.75);
\node at (4.1,1.95) {$t$};
\draw[Emerald,gluon] (2.5,1.0)--(3.0,2.5);
\node at (3.1,2.8) {$\color{Emerald} g$};
\draw (2.5,1.0)--(3.8,0.25);
\node at (3.1,0.35) {${\tilde S}$};
\draw (3.8,0.25)--(4.8,-0.5);
\node at (5.0,-0.75) {${\tilde G}$ (soft)};
\draw[dashed] (3.8,0.25)--(4.8,0.75);
\node at (4.3,0.7) {$S$};
\draw[electron] (5.8,1.5)--(4.8,0.75);
\draw[electron] (4.8,0.75)--(5.8,0.0);
\node at (6.25,1.5) {$b~{\color{Emerald} (g)}$};
\node at (6.25,0.0) {${\overline b}~{\color{Emerald} (g)}$};
\end{scope}
\end{tikzpicture}
\end{center}
\caption{The ${\tilde g} \to {\tilde t}_R \to {\tilde S} \to {\tilde G}$ simplified model. Left: diagram of decays.The green ``$\color{Emerald} +g$'' in the stop decay applies only to the $SY{\overline Y}$ scenario, not the $SH_uH_d$ one. Right: Feynman diagram for the most common decay chain. We show the $SH_u H_d$ scenario in black, with the green gluons indicating the most common decays in the alternative $SY{\overline Y}$ scenarios.
} \label{fig:simplifiedGluinoStopSinglino}
\end{figure}

The decay chain with a right-handed stop as the lightest MSSM superpartner splits into two cases: in the $SH_u H_d$ model, the stop decays as ${\tilde t} \to t {\tilde S}$, while in the $SY {\overline Y}$ model it decays as ${\tilde t} \to t g {\tilde S}$. Thus there are slightly different signatures in terms of kinematics and jet multiplicity, and we  provide two different exclusion plots. In both cases, we consider a gluino that decays to ${\overline t}{\tilde t}_R$ (and its charge conjugate) or, if phase space is insufficient, to the 3-body final state ${\overline b} W^- {\tilde t}_R$.

\begin{figure}[!h]\begin{center}
\includegraphics[width=0.45\textwidth]{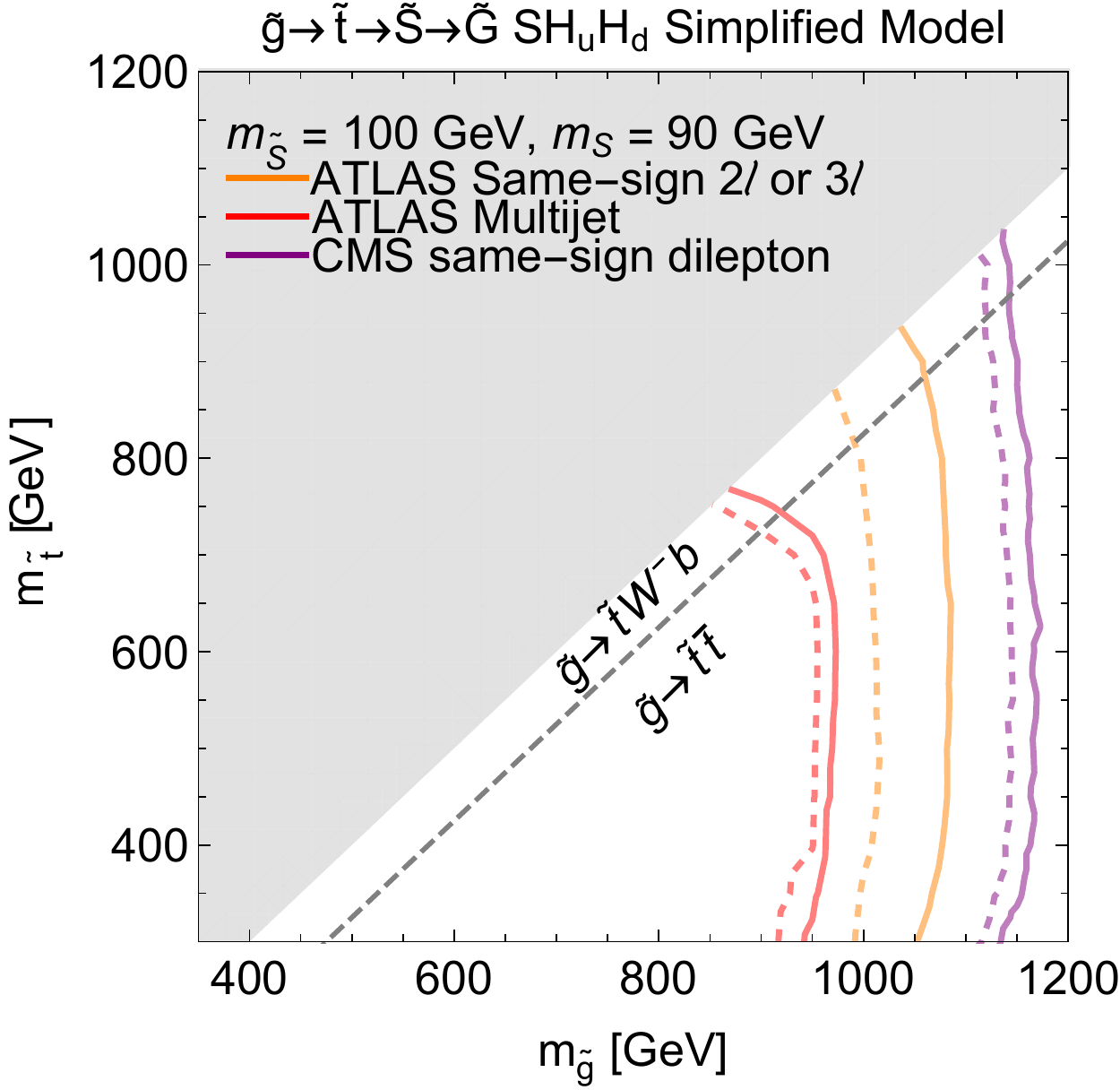} \quad \includegraphics[width=0.425\textwidth]{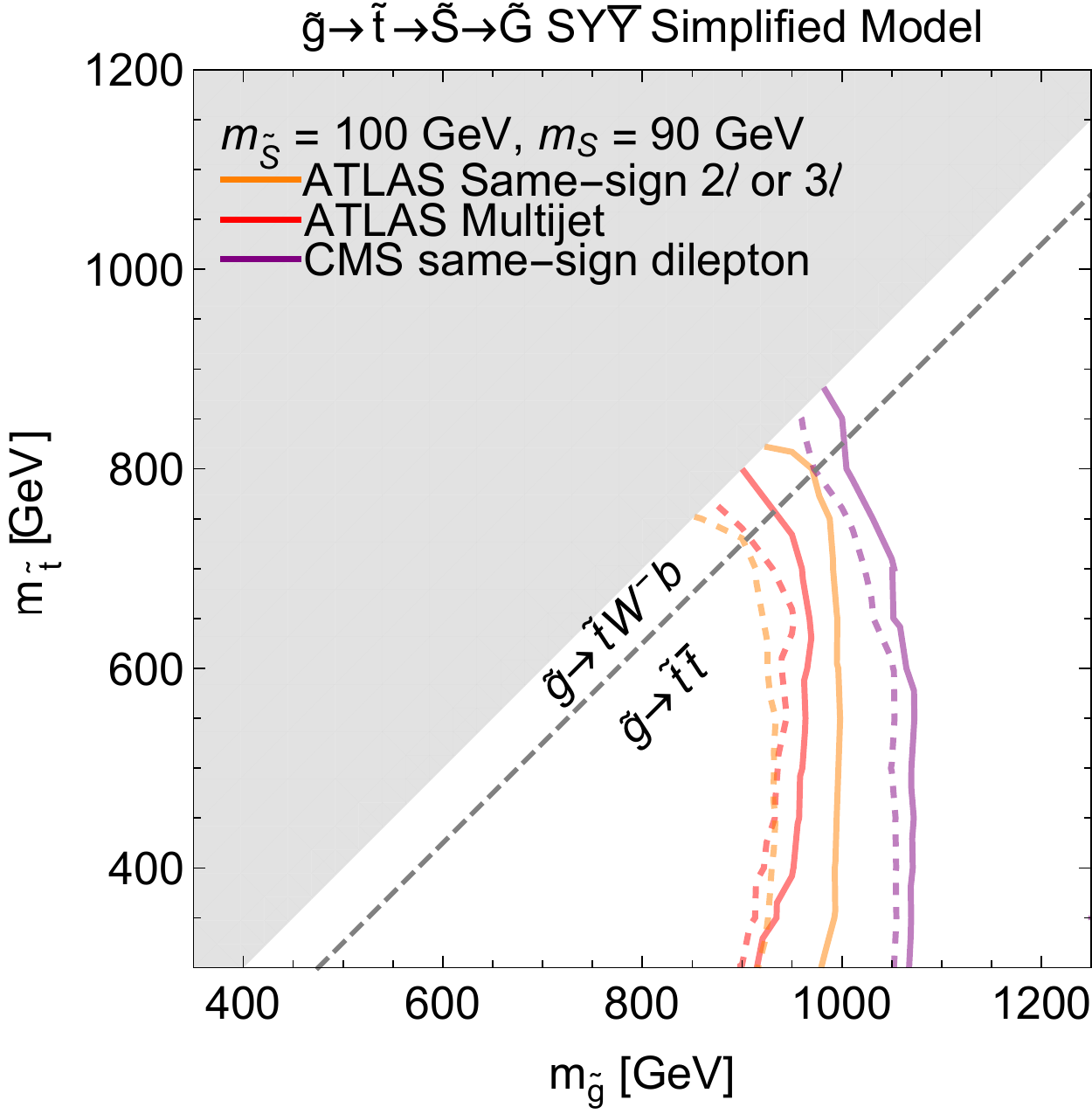}
\end{center}
\caption{Bound on the ${\tilde g} \to {\tilde t} \to {\tilde S} \to {\tilde G}$ decay chain. {\bf At left}: bound for the $SH_uH_d$ model. {\bf At right}: bound for the $SY{\overline Y}$ model. In both cases, solid lines correspond to 95\% $CL_s$ exclusion limits using the best signal region from a given search. The solid purple line is the bound from the CMS same-sign dilepton search with $b$-jets \cite{Chatrchyan:2013fea}. The solid orange line is the bound from the ATLAS same-sign 2$l$ or 3$l$ search~\cite{Aad:2014pda}. The solid red line is for the ATLAS multijet search~\cite{TheATLAScollaboration:2013xia,Aad:2015lea}. The dashed lines correspond to weakening the bounds by a factor of two (accounting for a possible overestimation of the search efficiency).}
\label{fig:boundGlStS}
\end{figure}%

Both models are well constrained by the ATLAS multijet search~\cite{TheATLAScollaboration:2013xia,Aad:2015lea}, the ATLAS same-sign dilepton or 3-lepton search~\cite{Aad:2014pda}, and the CMS same-sign dilepton search with $b$-jets \cite{Chatrchyan:2013fea}. The ATLAS multijet search signature is events with high jet multiplicities without any constraints on missing transverse momentum. Such searches are well suited for these models due to the stop decay, which produces events with jets and multiple leptons. Of the signal regions in the ATLAS same-sign dilepton search, the SR3b region places the tightest exclusion limits. This signal region includes events which have same-sign leptons or three leptons and at least five jets, of which at least three must be $b$-tagged. The strongest constraint that we find is from the CMS same-sign dilepton search.  The CMS search relies on two same-sign leptons above 20 GeV in $p_T$, multiple jets of which at least two are $b$-tagged, large $H_T > 400$ GeV, and moderate missing transverse momentum. As was already observed in ref~\cite{Evans:2013jna}, it is difficult to hide a gluino due to the large multiplicity of energetic particles in the final state.

\subsection{The ${\tilde g} \to {\tilde t} \to {\tilde H} \to {\tilde S} \to {\tilde G}$ Decay Chain}

\begin{figure}[!b]\begin{center}
\includegraphics[width=0.4\textwidth]{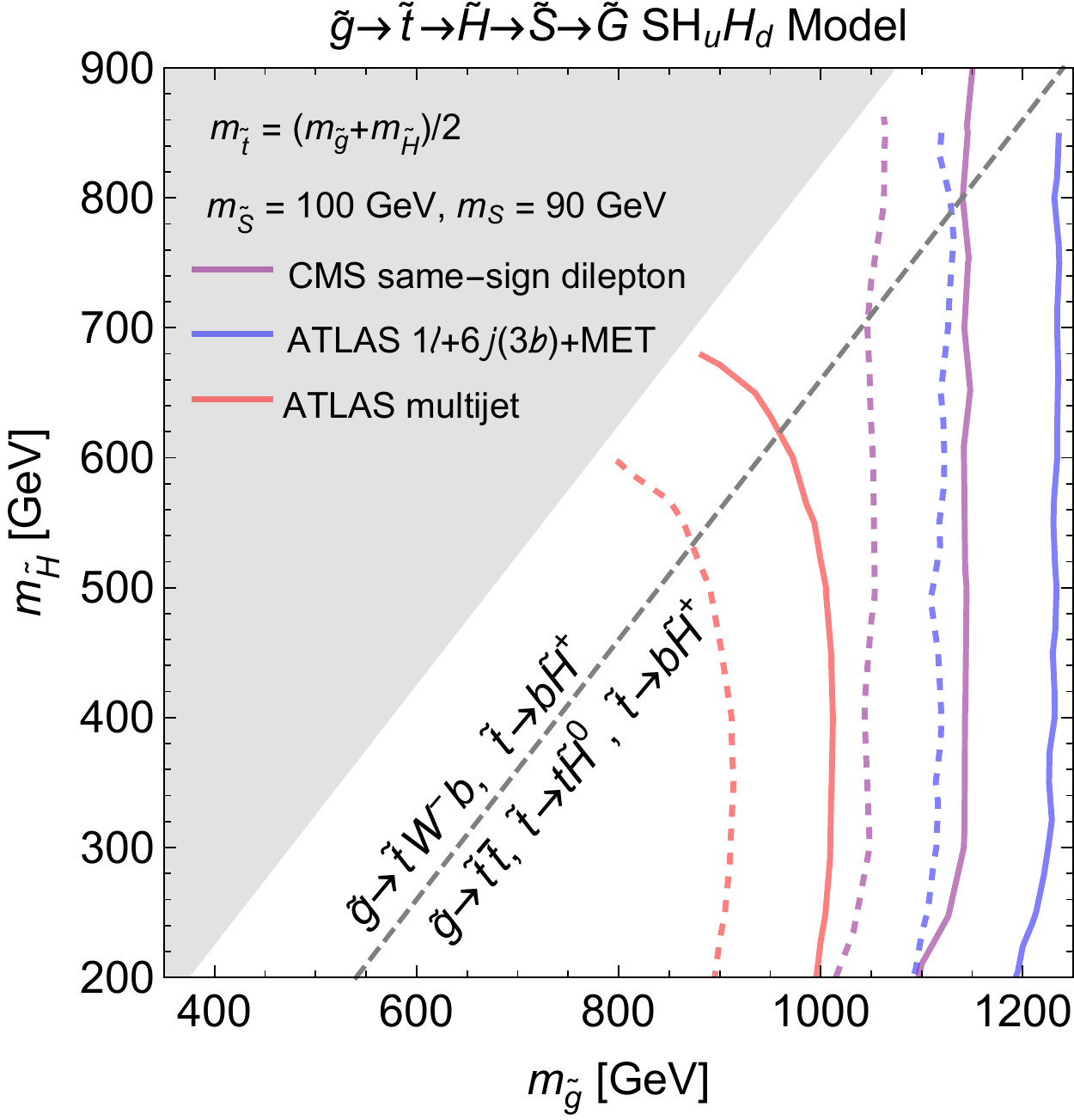} \quad \includegraphics[width=0.4\textwidth]{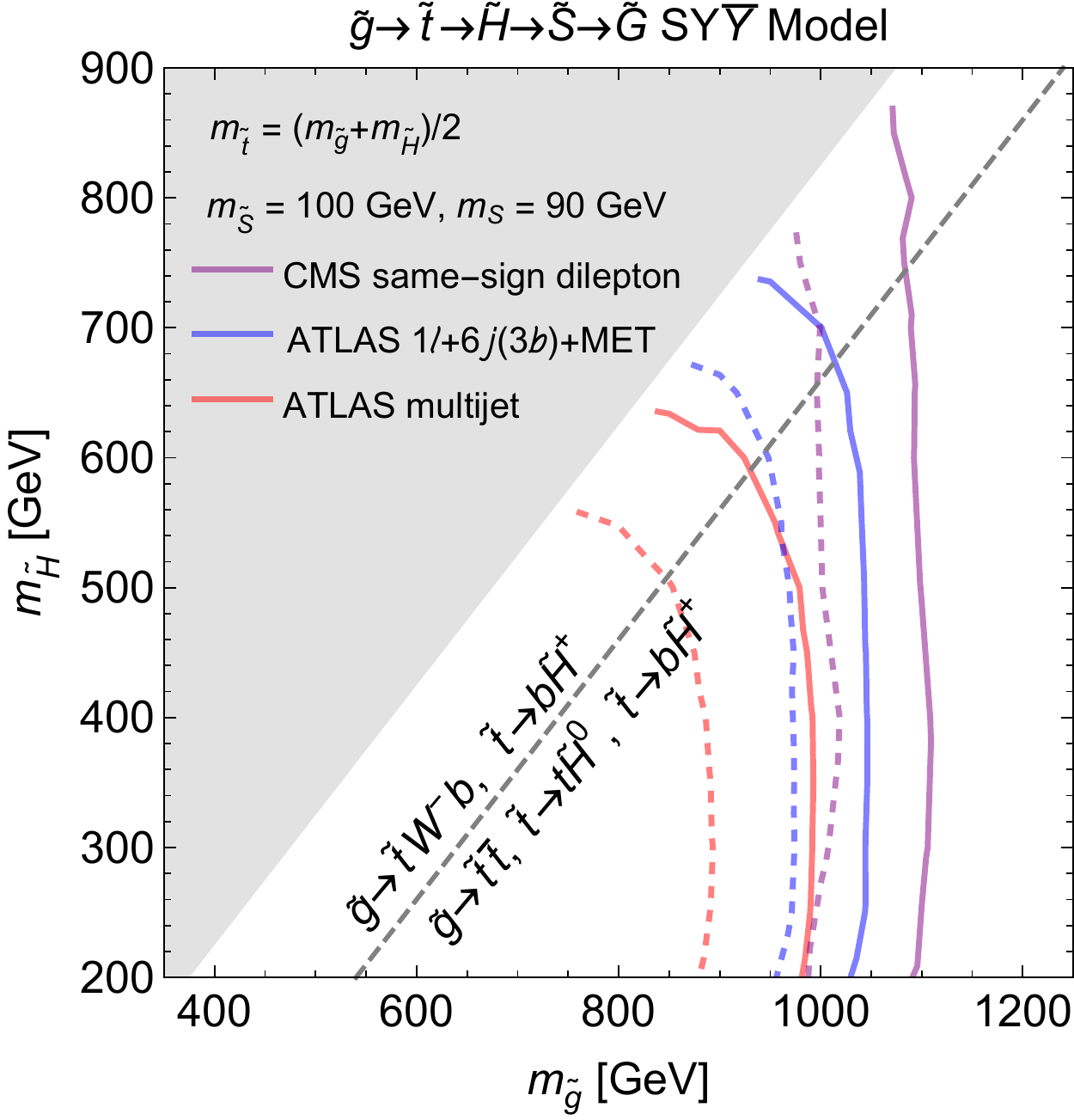}
\end{center}
\caption{Bound on the ${\tilde g} \to {\tilde t} \to {\tilde H} \to {\tilde S} \to {\tilde G}$ decay chain. {\bf At left:} bound for the $SH_uH_d$ model. {\bf At right:} bound for the $SY{\overline Y}$ model (right). The purple curves correspond to the CMS same-sign dilepton plus jets search  \cite{Chatrchyan:2013fea}; the red curves correspond to the ATLAS multijet search~\cite{TheATLAScollaboration:2013xia,Aad:2015lea} and the blue curves correspond to the ATLAS search requiring at least three $b$-jets plus missing energy~\cite{TheATLAScollaboration:2013tha,Aad:2014lra}. Solid lines correspond to 95\% $CL_s$ exclusion limits using the best signal region from a given search, and dashed lines weaken the bound by a factor of 2.}
\label{fig:gluinostopHiggsinobound}
\end{figure}

The longest decay chain produces large numbers of final state particles. Because of the top decay and the electroweak bosons emitted in Higgsino decays, there is ample opportunity for multiple ($b$-)jets and multiple leptons (including neutrinos leading to missing energy) to be present. As a result, although this decay chain looks different from those appearing in standard SUSY searches, it is quite easy to constrain and the bounds from existing data are already good. For instance, the CMS same-sign dilepton plus jets search~\cite{Chatrchyan:2013fea}, the ATLAS multijet search~\cite{TheATLAScollaboration:2013xia,Aad:2015lea} and the ATLAS search requiring at least three $b$-jets plus missing energy~\cite{TheATLAScollaboration:2013tha,Aad:2014lra} all exclude the gluino up to (1--1.1) TeV for the whole mass range of the Higgsino as shown in Figure~\ref{fig:gluinostopHiggsinobound}. Again this is consistent with the observation in ref~\cite{Evans:2013jna}.

\begin{figure}[H]
\begin{center}
\resizebox{0.85\textwidth}{!}{%
\begin{tikzpicture}[line width=1.5 pt]
\node at (1,5.3) {\Large ${\tilde g}$};
\draw[violet] (0,5.0)--(2,5.0);
\node at (1,4.0) {\Large ${\tilde t}_R$};
\draw[violet,dashed] (0,3.7)--(2,3.7);
\draw[-latex] (2.1,5.0) to[out=30,in=2,looseness=1.5] (2.1,3.7);
\node at (3.0,4.5) {$+t$};
\draw[-latex,dashed] (-0.1,3.7) to[out=210,in=140,looseness=1.5] (-0.1,2.4);
\draw[-latex,dashed] (2.1,3.7) to[out=30,in=2,looseness=1.5] (2.1,2.0);
\node at (-1.2, 3.0) {$+t$};
\node at (3.2, 3.0) {$+b$};
\draw[-latex,dashed] (-0.1,3.7) to[out=210,in=140,looseness=1.5] (-0.1,1.6);
\draw[violet] (0,1.6)--(2,1.6);
\draw[violet] (0,2.0)--(0.6,2.0);\draw[violet] (1.4,2.0)--(2,2.0);
\draw[violet] (0,2.4)--(2,2.4);
\node at (1.0,2.75) {\Large ${\tilde H}^0_2$};
\node at (1.0,1.25) {\Large ${\tilde H}^0_1$};
\node at (1.0,2.0) {\Large ${\tilde H}^\pm$};
\draw[violet] (0,0)--(2,0);
\draw[dashed] (0,-0.4)--(2,-0.4);
\draw[violet] (0,-2.3)--(2,-2.3);
\node at (1,0.3) {\Large ${\tilde S}$};
\node at (1,-0.7) {\Large $S$};
\node at (1,-2.0) {\Large ${\tilde G}~{\rm or}~{\tilde a}$};
\draw[-latex] (-0.1,0) to[out=210,in=140,looseness=1.5] (-0.1,-2.2);
\draw[-latex] (2.1,2.0) to[out=340,in=30,looseness=2] (2.1,0.1);
\node at (-2.3,0.75) {$+h,Z,S$};
\node at (3.7,1.0) {$+W^\pm$};
\draw[-latex] (-0.1,1.6) to[out=200,in=160,looseness=2] (-0.1,0.1);
\draw[-latex] (-0.1,2.4) to[out=200,in=170,looseness=2] (-0.3,0.1);
\node at (-1.4,-0.95) {$+S$};
\begin{scope}[shift={(5.2,1.5)}]
\draw (0,0)--(1.5,0);\draw[gluon] (0,0)--(1.5,0);
\node at (0.76,0.5) {${\tilde g}$};
\draw[scalar] (1.5,0)--(2.5,1.0);
\node at (1.95,0.85) {${\tilde t}_R$};
\draw[electron] (2.5,-1.0)--(1.5,0);
\node at (2.7,-1.2) {${\overline t}$};
\draw[electron] (2.5,1.0)--(3.8,1.85);
\node at (4.1,2.05) {$b$};
\draw (2.5,1.0)--(3.8,0.25);
\node at (3.1,0.35) {${\tilde H}^+$};
\draw[photon] (3.8,0.25)--(4.8,1.15);
\node at (5.15,1.2) {$W^+$};
\draw (3.8,0.25)--(4.8,-0.75);
\node at (4.2,-0.6) {${\tilde S}$};
\begin{scope}[shift={(1.0,-1.0)}]
\draw (3.8,0.25)--(4.8,-0.5);
\node at (5.0,-0.75) {${\tilde G}$ (soft)};
\draw[dashed] (3.8,0.25)--(4.8,0.75);
\node at (4.3,0.7) {$S$};
\draw[electron] (5.8,1.5)--(4.8,0.75);
\draw[electron] (4.8,0.75)--(5.8,0.0);
\node at (6.25,1.5) {$b~{\color{Emerald} (g)}$};
\node at (6.25,0.0) {${\overline b}~{\color{Emerald} (g)}$};
\end{scope}
\end{scope}
\end{tikzpicture}
}
\end{center}
\caption{The ${\tilde g} \to {\tilde t}_R \to {\tilde H} \to {\tilde S} \to {\tilde G}$ simplified model. 
} \label{fig:simplifiedGluinoStopHiggsinoSinglino}
\end{figure}
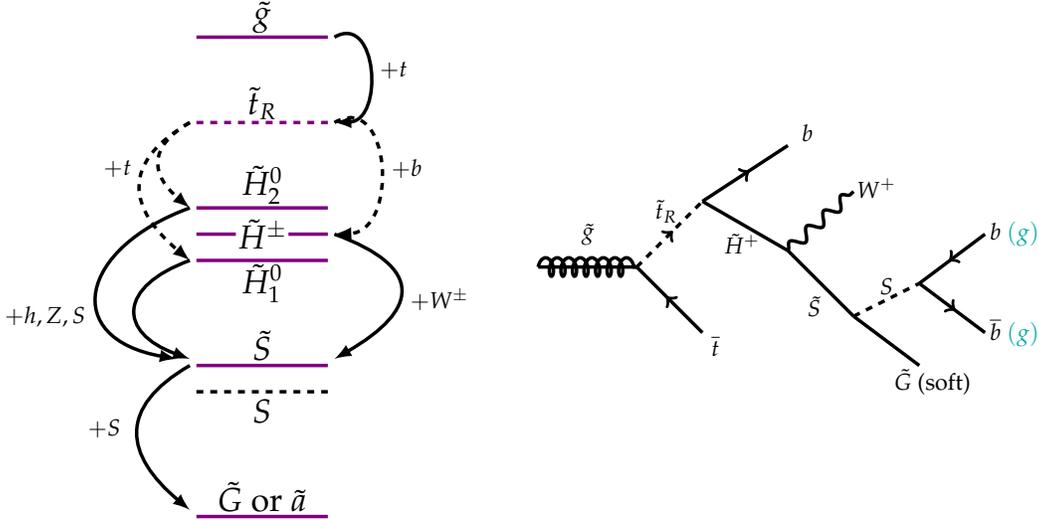

\section{Constraints on Stops}
\label{sec:stopconstraint}
\afterpage{\clearpage}

We now consider simplified models describing stop production, which are relevant when the gluino is somewhat heavier than the stop.  In certain cases, the gluino may be naturally made much heavier, such as in models with Dirac gauginos~\cite{Fox:2002bu, Kribs:2012gx, Alves:2015kia, Alves:2015bba}.  In Section~\ref{subsec:stopsinglino} we study topologies in which the stop decays directly to the singlino; in Section~\ref{subsec:stopHiggsino}, the stop decays through an intermediate Higgsino.

\subsection{The ${\tilde t} \to {\tilde S} \to {\tilde G}$ Decay Chain}
\label{subsec:stopsinglino}

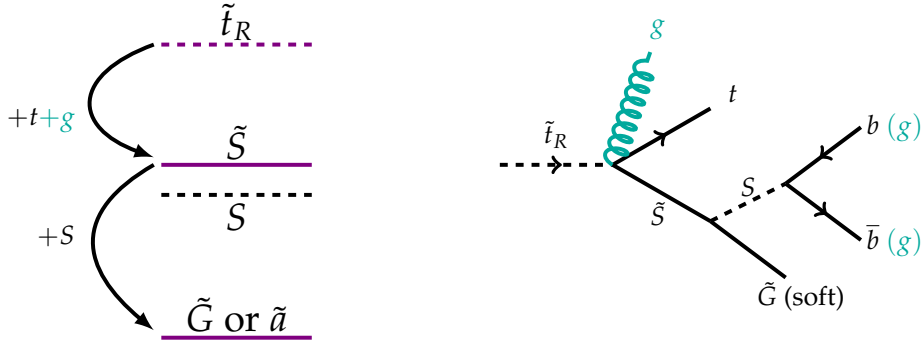
\begin{figure}[!t]
\begin{center}
\begin{tikzpicture}[line width=1.5 pt]
\draw[dashed,violet] (0,1.6)--(2,1.6);
\node at (1,1.9) {\Large ${\tilde t}_R$};
\draw[violet] (0,0)--(2,0);
\draw[dashed] (0,-0.4)--(2,-0.4);
\draw[violet] (0,-2.3)--(2,-2.3);
\node at (1,0.3) {\Large ${\tilde S}$};
\node at (1,-0.7) {\Large $S$};
\node at (1,-2.0) {\Large ${\tilde G}~{\rm or}~{\tilde a}$};
\draw[-latex] (-0.1,0) to[out=210,in=140,looseness=1.5] (-0.1,-2.2);
\node at (-1.6,0.6) {$+t{\color{Emerald}+g}$};
\draw[-latex] (-0.1,1.6) to[out=200,in=160,looseness=2] (-0.1,0.1);
\node at (-1.4,-0.95) {$+S$};
\begin{scope}[shift={(3.5,-1.0)}]
\draw[scalar] (1.0,1.0)--(2.5,1.0);
\node at (1.75,1.4) {${\tilde t}_R$};
\draw[electron] (2.5,1.0)--(3.8,1.75);
\node at (4.1,1.95) {$t$};
\draw[Emerald,gluon] (2.5,1.0)--(3.0,2.5);
\node at (3.1,2.8) {$\color{Emerald} g$};
\draw (2.5,1.0)--(3.8,0.25);
\node at (3.1,0.35) {${\tilde S}$};
\draw (3.8,0.25)--(4.8,-0.5);
\node at (5.0,-0.75) {${\tilde G}$ (soft)};
\draw[dashed] (3.8,0.25)--(4.8,0.75);
\node at (4.3,0.7) {$S$};
\draw[electron] (5.8,1.5)--(4.8,0.75);
\draw[electron] (4.8,0.75)--(5.8,0.0);
\node at (6.25,1.5) {$b~{\color{Emerald} (g)}$};
\node at (6.25,0.0) {${\overline b}~{\color{Emerald} (g)}$};
\end{scope}
\end{tikzpicture}
\end{center}
\caption{The ${\tilde t}_R \to {\tilde S} \to {\tilde G}$ simplified model. Left: diagram of decays.The green ``$\color{Emerald} +g$'' in the stop decay applies only to the $SY{\overline Y}$ scenario, not the $SH_uH_d$ one. Right: Feynman diagram for the most common decay chain. We show the $SH_u H_d$ scenario in black, with the green gluons indicating the most common decays in the alternative $SY{\overline Y}$ scenarios.
} \label{fig:simplifiedStopSinglinoS}
\end{figure}

As described in Section~\ref{subsec:gluinostopsinglino}, the stop has two relevant decay modes when it is lighter than the Higgsino.  In models without significant singlet-Higgs mixing, such as $S Y \bar{Y}$, the stop decays through an off-shell gluino: $\tilde{t} \rightarrow g t \tilde{S}$.  When the singlet and the Higgs do mix, as in the $S H_u H_d$ model, the stop decays through its Yukawa coupling directly to $t \tilde{S}$.  These two cases are further distinguished by the decay modes of the singlet, which is itself produced in decays of the singlino.  In the $S H_u H_d$ model, mixing causes the singlet to inherit the branching fractions of the Higgs, so that it dominantly decays to $b \bar{b}$.  Coupling to vector-like quarks instead ($S Y \bar{Y}$) gives a dominant decay channel to gluons.  These decays are shown schematically in Figure~\ref{fig:simplifiedStopSinglinoS}. 

\begin{figure}[!h]\begin{center}
\includegraphics[width=\textwidth]{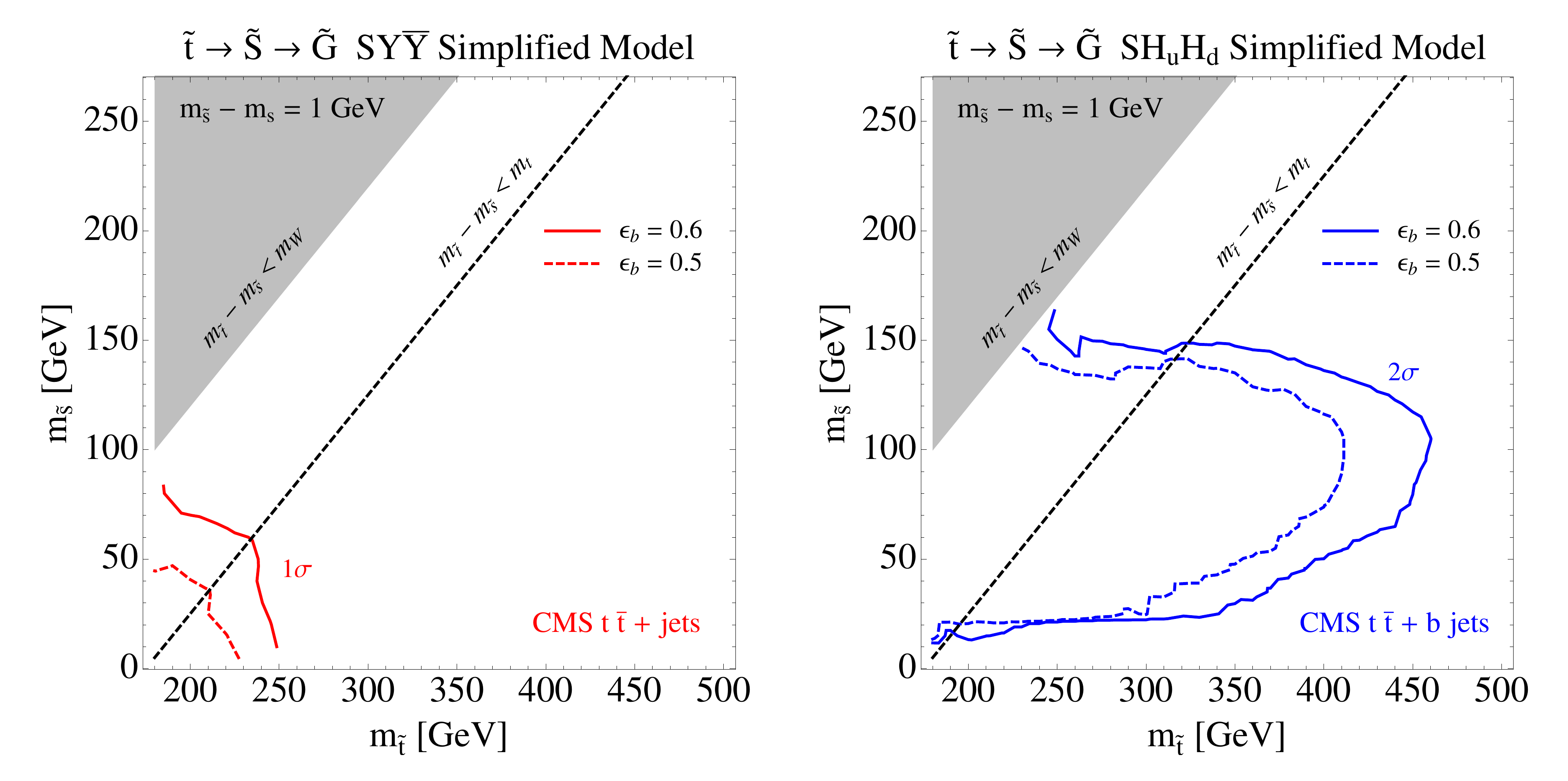}
\end{center}
\caption{Bounds on the ${\tilde t} \to {\tilde S} \to {\tilde G}$ decay chain. {\bf At left:} 68\% $CL_s$ exclusion bound on the $SY{\overline Y}$ model from the CMS measurement of $t\bar{t}$ + jets~\cite{Khachatryan:2015oqa}.   {\bf At right:} 95\% $CL_s$ exclusion bound on the $SH_uH_d$ model from the $t\bar{t}$ + multi-$b$-jet distribution.  Solid curves correspond to a nominal $60\%$ $b$-tagging efficiency, while dashed curves reflect a conservative rescaling to $50\%$ as a crude representation of systematic uncertainty in our detector simulation.  To the left of the dashed black line, tops produced in stop decays are off-shell.  Regions of parameter space with off-shell W bosons are shaded in grey.}
\label{fig:stopjetslimits}
\end{figure}

The similarity of the signal to $t \bar{t} + $ jets production in QCD hides these topologies from traditional SUSY searches, motivating precision study of the Standard Model process.  Both ATLAS and CMS have performed measurements of the differential $t\bar{t}$ + jets production cross section at 8~\cite{Khachatryan:2015oqa, Aad:2015yja, Khachatryan:2015mva} and 13 TeV~\cite{ATLAS-CONF-2015-065, CMS-PAS-TOP-15-013}.  In order to estimate the sensitivity of these measurements to the presence of stop decays, we used simulation to reproduce jet multiplicity distributions in $t\bar{t}$ events with semi-leptonic top decays measured in~\cite{Khachatryan:2015oqa}.  As these measurements have comparable systematic uncertainties, we chose to focus on one to illustrate the sensitivity to new physics.  We validated the (b-)jet multiplicity distributions on simulated $t\bar{t}$ events, as described in Appendix~\ref{app:ttbarplusjets}, before using them to constrain tops produced in stop decays.  These limits are shown in Figure~\ref{fig:stopjetslimits}.  Because the correlation of systematic uncertainties between bins is not presented, we set a limit using the single bin with $N_{\text{jets}} > 4$ or $N_{b\text{-jets}} > 2$ with the highest expected sensitivity.  If these correlations are provided, future measurements may have improved sensitivity.

In the $S H_u H_d$ model, shown on the right-hand side of the figure, the majority of events contain 6 $b$-jets.  Accordingly, the bin with the highest number of observed $b$-jets provides the most sensitivity, constraining stops to be heavier than $\sim 400$~GeV for singlino masses near $100$~GeV at the $95\%$ confidence level.  As the singlet mass is raised, the branching fraction of $S$ to $b\bar{b}$ decreases rapidly, reducing the sensitivity accordingly.  Above $\sim 150$~GeV, the majority of decays go to $W W^*$.  Aside from reducing the multiplicity of $b$-jets in the event, leptonically-decaying $W$ bosons produce missing energy, spoiling the stealth mechanism.  As a result, we do not focus on this region of parameter space.  Traditional SUSY search strategies looking for combinations of $b$-jets, multi-leptons, and missing energy should be able to effectively probe this region.

For the $SY\bar{Y}$ simplified model, shown on the left-hand side of the figure, the only $b$-jets come from top decays.  The majority of events contain 8 jets, of which 6 are light-flavor, so that they populate the tails of the (un-tagged) jet multiplicity distribution.  Large uncertainties in the $t\bar{t}$ cross section at large jet multiplicity significantly reduce the sensitivity, and we find that these stops are completely unconstrained at the 95\% confidence level.  The bin with $N_{\text{jets}} = 5$ has the smallest uncertainties, providing a limit on stop masses below $\sim 250$~GeV at the 68\% confidence level.  Thus, despite the absence of Higgsinos from the bottom of the spectrum, the stealthy nature of stop decays in this simplified model allows the Higgsino mass to remain at or below several hundred GeV, within the natural range of the MSSM parameter space.

In addition to decays with large numbers of $(b-)$jets, both the $S H_u H_d$ and $S Y \bar{Y}$ models feature a sub-dominant decay mode of the singlet to diphotons, with a branching fraction of $\mathcal{O} (10^{-3} )$.  CMS conducted a search for stops decaying through Higgsinos to gravitinos and Higgses~\cite{Chatrchyan:2013mya}, selecting events with $b$-jets and Higgs decays to photons.  We recast the event selection, floating the invariant mass cut on the photons to match the singlet mass at each point.  We then used the provided diphoton invariant mass distributions to set a conservative limit on decays involving photons, requiring that the signal not produce more events than the total observed at the 95\% confidence level.  Although this search showed promising sensitivity at somewhat larger photon branching ratios, we were unable to set a limit in either simplified model from decays to photons.  Note that the reach of this search is currently limited by statistical uncertainties, so that higher luminosity at Run 2 should lead to improved sensitivity.

Both search strategies, investigating $t\bar{t}$ production at large jet multiplicity and searching for resonant diphoton production in association with $b-$jets, showed promising, though marginal, sensitivity to stop production.  In the case of diphotons, it seems plausible that a dedicated search outside of the Higgs mass window with a less conservative limit-setting procedure may be able to probe these topologies.  More thorough study of $t\bar{t}$ production at high jet multiplicity, perhaps combining with kinematic variables to improve sensitivity, is also an interesting future direction.

\subsection{The ${\tilde t} \to {\tilde H} \to {\tilde S} \to {\tilde G}$ Decay Chain}
\label{subsec:stopHiggsino}

\begin{figure}[H]
\begin{center}
\begin{tikzpicture}[line width=1.5 pt]
\node at (1,4.0) {\Large ${\tilde t}_R$};
\draw[violet,dashed] (0,3.7)--(2,3.7);
\draw[-latex,dashed] (-0.1,3.7) to[out=210,in=140,looseness=1.5] (-0.1,2.4);
\draw[-latex,dashed] (2.1,3.7) to[out=30,in=2,looseness=1.5] (2.1,2.0);
\node at (-1.2, 3.0) {$+t$};
\node at (3.2, 3.0) {$+b$};
\draw[-latex,dashed] (-0.1,3.7) to[out=210,in=140,looseness=1.5] (-0.1,1.6);
\draw[violet] (0,1.6)--(2,1.6);
\draw[violet] (0,2.0)--(0.6,2.0);\draw[violet] (1.4,2.0)--(2,2.0);
\draw[violet] (0,2.4)--(2,2.4);
\node at (1.0,2.75) {\Large ${\tilde H}^0_2$};
\node at (1.0,1.25) {\Large ${\tilde H}^0_1$};
\node at (1.0,2.0) {\Large ${\tilde H}^\pm$};
\draw[violet] (0,0)--(2,0);
\draw[dashed] (0,-0.4)--(2,-0.4);
\draw[violet] (0,-2.3)--(2,-2.3);
\node at (1,0.3) {\Large ${\tilde S}$};
\node at (1,-0.7) {\Large $S$};
\node at (1,-2.0) {\Large ${\tilde G}~{\rm or}~{\tilde a}$};
\draw[-latex] (-0.1,0) to[out=210,in=140,looseness=1.5] (-0.1,-2.2);
\draw[-latex] (2.1,2.0) to[out=340,in=30,looseness=2] (2.1,0.1);
\node at (-2.4,0.75) {$+h,Z,S$};
\node at (3.7,1.0) {$+W^\pm$};
\draw[-latex] (-0.1,1.6) to[out=200,in=160,looseness=2] (-0.1,0.1);
\draw[-latex] (-0.1,2.4) to[out=200,in=170,looseness=2] (-0.3,0.1);
\node at (-1.4,-0.95) {$+S$};
\begin{scope}[shift={(4.9,0.8)}]
\draw[scalar] (1.25,1.0)--(2.5,1.0);
\node at (1.05,1.0) {${\tilde t}_R$};
\draw[electron] (2.5,1.0)--(3.5,2.0);
\node at (3.7,2.15) {$b~\color{orange}(t)$};
\draw (2.5,1.0)--(3.8,0.25);
\node at (3.1,0.35) {${\tilde H}^+$};
\node at (3.1,-0.15) {$\color{orange} ({\tilde H}^0_{1,2})$};
\draw[photon] (3.8,0.25)--(4.8,1.15);
\node at (5.25,1.35) {$W^+~{\color{orange} (Z^0, h^0, S)}$};
\draw (3.8,0.25)--(4.8,-0.75);
\node at (4.2,-0.6) {${\tilde S}$};
\begin{scope}[shift={(1.0,-1.0)}]
\draw (3.8,0.25)--(4.8,-0.5);
\node at (5.0,-0.75) {${\tilde G}$ (soft)};
\draw[dashed] (3.8,0.25)--(4.8,0.75);
\node at (4.3,0.7) {$S$};
\draw[electron] (5.8,1.5)--(4.8,0.75);
\draw[electron] (4.8,0.75)--(5.8,0.0);
\node at (6.25,1.5) {$b~{\color{Emerald} (g)}$};
\node at (6.25,0.0) {${\overline b}~{\color{Emerald} (g)}$};
\end{scope}
\end{scope}
\end{tikzpicture}
\end{center}
\caption{The ${\tilde t}_R \to {\tilde H} \to {\tilde S} \to {\tilde G}$ simplified model. Left: diagram of possible decays. Right: example Feynman diagram for the most common decay chain in the $SH_u H_d$ model (black), plus alternative singlet decay in the $SY{\overline Y}$ model (green) and other possible stop decay chains in both models (orange).
} \label{fig:simplifiedStopHiggsinoSinglino}
\end{figure}
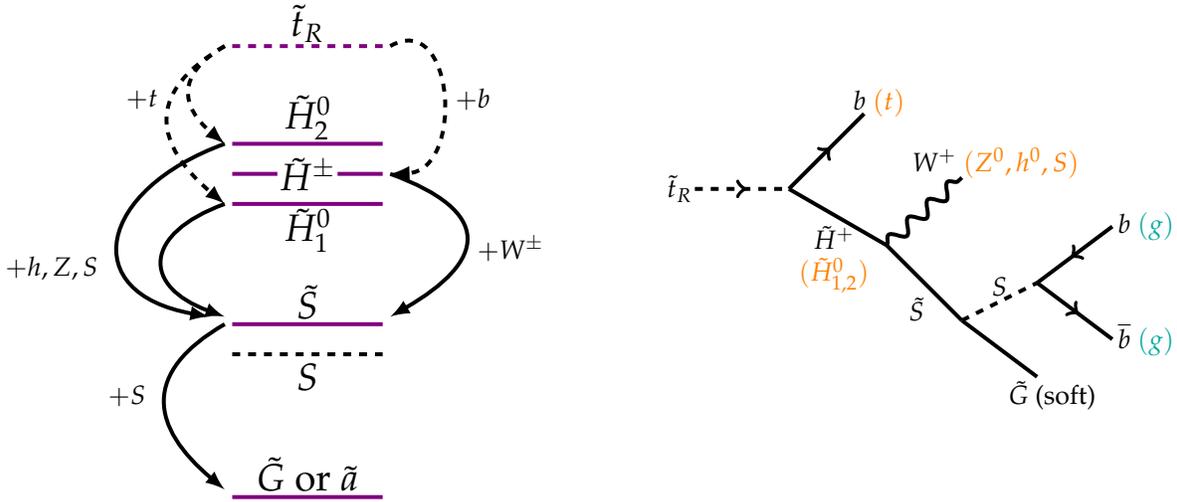

\begin{figure}[!h]\begin{center}
\includegraphics[width=0.45\textwidth]{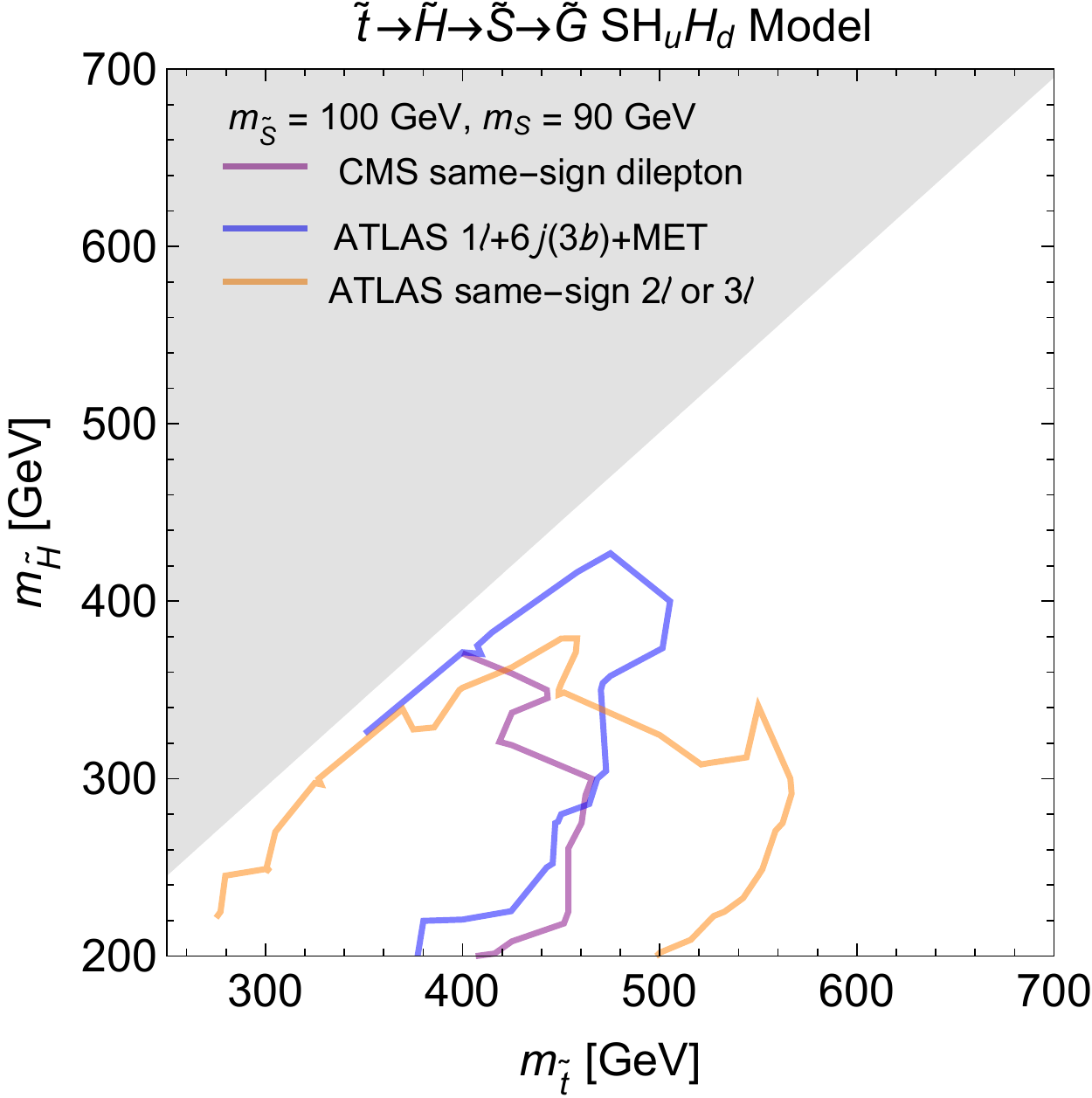} \quad \includegraphics[width=0.45\textwidth]{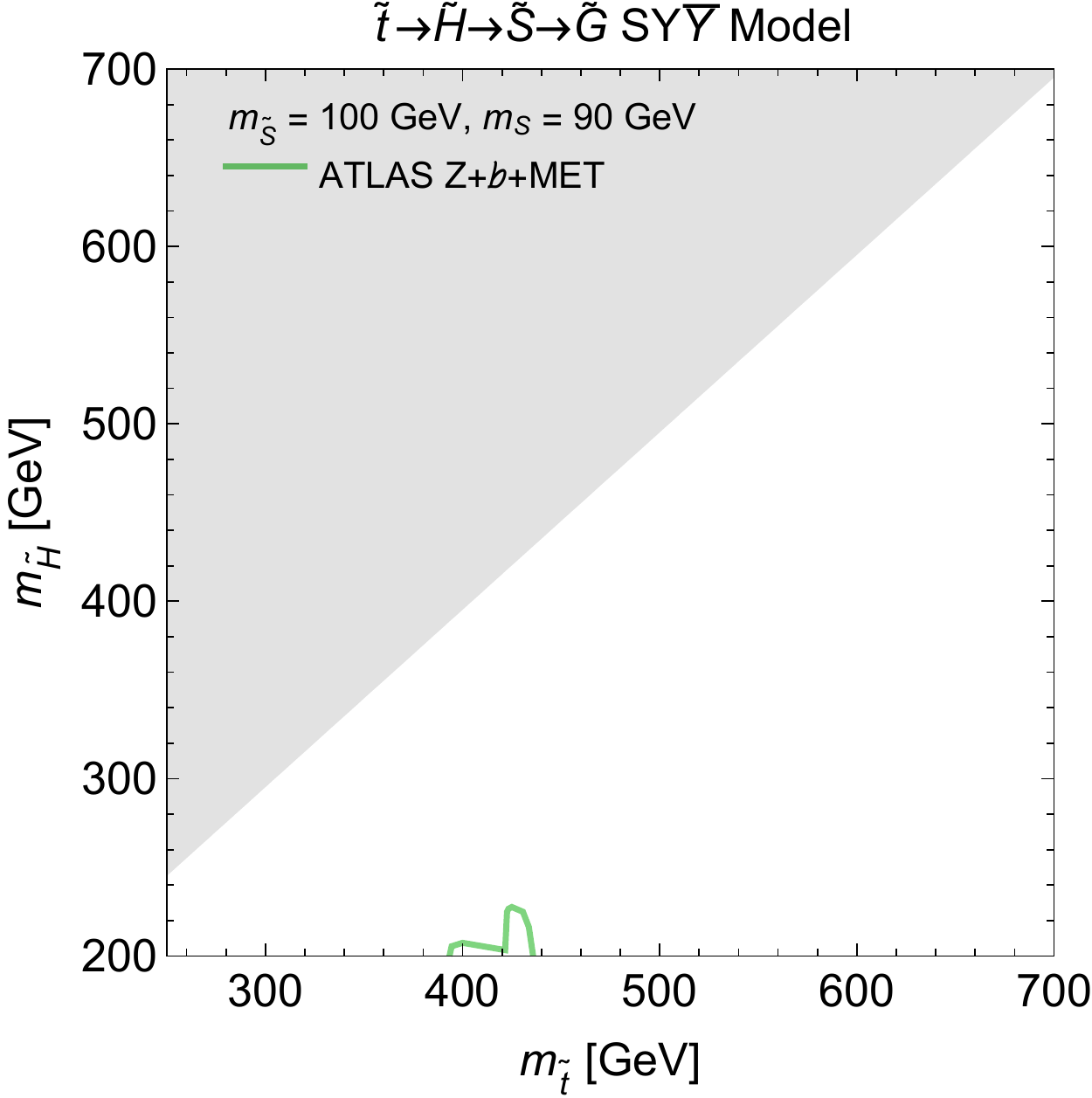}
\end{center}
\caption{Bound on the ${\tilde t} \to {\tilde H} \to {\tilde S} \to {\tilde G}$ decay chain for $SH_uH_d$ model (left) and $SY{\overline Y}$ model (right). {\bf Left}: The purple bound comes from the CMS same-sign dilepton plus $b$-jet search \cite{Chatrchyan:2013fea}; the orange line from the ATLAS same-sign dilepton or 3-lepton plus 3 $b$-jet channel \cite{Aad:2014pda}; and the blue line from the ATLAS 3 $b$-jet plus missing transverse momentum channel \cite{Aad:2014lra}. {\bf Right}: for the $SY{\overline Y}$, we only find a bound from the heavy stop search in the $Z+b+\met$ channel at ATLAS \cite{Aad:2014mha}. Unlike the other figures, we omit dashed lines because a factor of 2 uncertainty in efficiency leads to no bound at all.}
\label{fig:boundonStHiS}
\end{figure}

This decay chain will lead to more particles in the final state compared to the previous one. 
In the $SH_uH_d$ model, the longer stop decay chain will produce a lot of $b$ quarks in the final state. The stop will always decay to top plus neutralino or bottom plus chargino, giving a $b$-jet on each side of the event. The singlets decay dominantly to $b$ quarks and yield more $b$-jets. We found that searches requiring multiple $b$-jets associated with missing energy and/or multiple leptons could be sensitive to this type of simplified model. Such searches include, for example,  ATLAS searches in the final state with at least 3 $b$-jets plus missing energy~\cite{Aad:2014lra}, two same-sign leptons or three leptons with at least 3 $b$-jets~\cite{Aad:2014pda}, and CMS searches for same-sign dilepton plus at least 2 $b$-jets~\cite{Chatrchyan:2013fea}. The constraints from the first two searches are obtained using the CheckMATE package~\cite{Drees:2013wra}. 
The result is demonstrated in the left panel of Figure~\ref{fig:boundonStHiS}. Stops are excluded up to 500--550 GeV in the non-degenerate region. In the $SY{\overline Y}$ model, there are fewer $b$ quarks in the final state, as the singlet dominantly decays to gluons. We found a weak constraint from the search for  stop production in the $Z + b + \met$ channel~\cite{Aad:2014mha} as shown in the right panel of Figure~\ref{fig:boundonStHiS}. In this simplified model (both $SH_u H_d$ and $SY{\overline Y}$), assuming a conservative factor of 2 in efficiency makes the constraint disappear altogether.

\section{Outlook}
\label{sec:outlook}

We have presented a set of simplified models that highlight possible topologies generated by natural models of Stealth SUSY.  The simplified models in this paper generate different signals than simplified models motivated by the MSSM.  We have determined limits on our simplified models by recasting searches  from Run 1 of the LHC\@. 
If gluinos decay to top quarks, their mass is significantly constrained,  up to around one TeV, due to the high multiplicity of $W$, $Z$, and $b$ particles that are be generated in their decays. (A similar conclusion was emphasized in \cite{Evans:2013jna}.)  However, gluino decays to light flavor jets and no missing energy are much less constrained, and the gluino can be as light as $\sim 400-600$~GeV for certain values of the singlino mass.  We also find that bounds on stops are significantly weaker than in the MSSM.   In the case each stop decay to multiple $b$-jets, as in our $S H_u H_d$ model, stop limits extend to $\sim 400-500$~GeV.  In the case stop decays produce extra light flavor jets, as in the $S Y \bar Y$ model, we find no meaningful constraint from LHC searches.  This implies that Stealth SUSY stops are less constrained that stops in baryonic $R$-parity violating scenarios, where paired dijet searches are sensitive to $350-400$~GeV stops~\cite{Khachatryan:2014lpa}.  We note that our ${\tilde t} \to {\tilde S} \to {\tilde G}$ model produces signals that are similar to SM $t{\overline t}$ events with extra jets.  This simplified model could provide an interesting opportunity for applying recent progress in perturbative QCD toward better signal/background separation.

There are also several opportunities for future experimental searches to probe interesting Stealth SUSY topologies.  In order to facilitate this, we are posting tools for simulation of our simplified models on the web at \url{http://users.physics.harvard.edu/~mreece/stealthsusy/}.  Possibilities of interest include:
\begin{itemize}
\item Stop simplified models with singlet decays to $b\bar{b}$ can easily produce events with six or more $b$ quarks, motivating searches for events with very high $b$-jet multiplicity.
\item In the region of parameter space in which the stealth SUSY stop decays to an {\em off-shell} top plus additional jets, one may try to veto events with a reconstructed {\em on-shell} top to reduce the background.
\item While we have checked that an existing CMS search for a stop in the channel with $h \to \gamma \gamma$~\cite{Chatrchyan:2013mya} does not bound our simplified models, searching for diphoton resonances in the context of busy events may probe topologies in which the singlet scalar has a (small) branching fraction to photons.
\end{itemize}

The LHC has recently begun to explore the 13 TeV energy realm.  This new energy scale presents a remarkable opportunity to produce and discover new states.   In order to leverage the full power of the LHC, it is important to identify promising signatures that are both consistent with 8 TeV limits and discoverable at 13 TeV\@.  Stealth SUSY is an excellent example: the reduced missing energy and higher multiplicity final states, compared to the MSSM,  dramatically reduce the power of the 8 TeV LHC\@.  On the other hand, the natural parameter space of Stealth SUSY points to stops and gluinos within kinematic reach of the 13 TeV LHC\@.  The simplified models we have identified in this paper present promising signatures and topologies, which have the potential to soon greatly clarify the status of naturalness of the weak scale.

\section*{Acknowledgments}
We thank Kyle Cranmer, Raffaele D'Agnolo, Monica D'Onofrio, Yuri Gershtein, Eva Halkiadakis, Jim Hirschauer, Yuichiro Nakai, Ryosuke Sato, Matt Strassler, Jesse Thaler, and Marc Weinberg for useful conversations. RK is supported by the Department of Energy Office of Science Graduate Fellowship Program (DOE SCGF), made possible in part by the American Recovery and Reinvestment Act of 2009, administered by ORISE-ORAU under contract no.~DE-AC05-06OR23100. The work of MR is supported in part by the NSF Grant PHY-1415548. Some computations in this paper were run on the Odyssey cluster supported by the FAS Division of Science, Research Computing Group at Harvard University. This work was supported in part by the National Science Foundation under Grant No. PHYS-1066293 and the hospitality of the Aspen Center for Physics.

\appendix

\section{Details of Branching Ratios}
\label{app:branching}

Throughout this appendix we use the notation
\beq
q(x,y) \equiv 1 + x^2 + y^2 - 2 x - 2 y - 2 x y
\eeq
for the phase-space function often denoted $\lambda$ in the literature. (We reserve $\lambda$ for a coupling in our model.) If a particle of mass $M$ decays to particles of mass $m_1$ and $m_2$, the momentum of a daughter particle in the rest frame of the decaying particle is $\frac{1}{2}q(m_1^2/M^2, m_2^2/M^2)^{1/2}M$.

\subsection{Gluinos to squarks}

We expect gluinos to decay approximately democratically to all squark flavors, but there is some dependence on the quark and squark masses, both from the standard phase space factor and from the matrix element:
\beq
\Gamma({\tilde g} \to {\overline q} \tilde{q}) \propto q(m_q^2/m_{\tilde g}^2, m_{\tilde q}^2/m_{\tilde g}^2)^{1/2} \left(1 + \frac{m_q^2}{m_{\tilde g}^2} - \frac{m_{\tilde q}^2}{m_{\tilde g}^2}\right).
\eeq

\subsection{Stops to Higgsinos}

Stop decays can produce either a top quark or a $b$-quark: ${\tilde t} \to t \chi^0$ and ${\tilde t} \to b \chi^+$. For generic mixing between left- and right-handed stops, both decays can be important (see, e.g., ref.~\cite{Graesser:2012qy}). However, for the purpose of having few-parameter simplified models we focus dominantly on either pure left-handed or pure right-handed stops, and we wish to obtain an approximate simplified expression for the branching ratios. These decays arise from the superpotential terms
\beq
W_{\rm stop} = y_t Q_3 H_u u^c_3 + y_b Q_3 H_d d^c_3,
\eeq
with the left-handed stop living in $Q_3$ and the right-handed (anti-)stop in $u^c_3$. From this we can see that the left-handed stop decays to $t + \chi^0$ via the top Yukawa and to $b + \chi^+$ via the bottom Yukawa. The latter would be competitive only if the former is suppressed by phase space or if $\tan \beta$ is very large. However, in natural scenarios $\tan \beta$ is not expected to be very large in light of experimental constraints on the stop--Higgsino loop contribution to $b \to s\gamma$~\cite{Blum:2012ii,Katz:2014mba}. Similarly, the left-handed sbottom prefers the $t + \chi^-$ mode to $b + \chi^0$ unless phase space is small or $\tan \beta$ is very large. On the other hand, the right-handed stop can decay to both $t + \chi^0$ and $b + \chi^+$ via the top Yukawa, so we expect the two modes to be competitive.

To good approximation (neglecting the $b$-quark mass) the branching ratios of stops decaying to Higgsinos when there is available two-body phase space are:
\beq
{\rm If}~m_{\tilde t} > m_t + \mu: & & \nonumber \\
\hline
{\rm Br}({\tilde t}_L \to {\tilde H}^0_1 t_R) & \approx & 50\%, \nonumber \\
{\rm Br}({\tilde t}_L \to {\tilde H}^0_2 t_R) & \approx & 50\%. \\
\hline
{\rm Br}({\tilde b}_L \to {\tilde H}^- t_R) & \approx & 100\%. \\
\hline
{\rm Br}({\tilde t}_R \to {\tilde H}^0_1 t_L) & \approx & \frac{1}{2} \frac{q(m_t^2/m_{\tilde t}^2, \mu^2/m_{\tilde t}^2)^{1/2} \left(m_{\tilde t}^2 - m_t^2 - \mu^2\right)}{q(0, \mu^2/m_{\tilde t}^2)^{1/2} \left(m_{\tilde t}^2 - \mu^2\right) + q(m_t^2/m_{\tilde t}^2, \mu^2/m_{\tilde t}^2)^{1/2} \left(m_{\tilde t}^2 - m_t^2 - \mu^2\right)} \to 25\%, \nonumber \\
{\rm Br}({\tilde t}_R \to {\tilde H}^0_2 t_L) & \approx &  \frac{1}{2} \frac{q(m_t^2/m_{\tilde t}^2, \mu^2/m_{\tilde t}^2)^{1/2} \left(m_{\tilde t}^2 - m_t^2 - \mu^2\right)}{q(0, \mu^2/m_{\tilde t}^2)^{1/2} \left(m_{\tilde t}^2 - \mu^2\right) + q(m_t^2/m_{\tilde t}^2, \mu^2/m_{\tilde t}^2)^{1/2} \left(m_{\tilde t}^2 - m_t^2 - \mu^2\right)} \to 25\%, \nonumber \\
{\rm Br}({\tilde t}_R \to {\tilde H}^+ b_L) & \approx & \frac{q(0, \mu^2/m_{\tilde t}^2)^{1/2} \left(m_{\tilde t}^2  - \mu^2\right)}{q(0, \mu^2/m_{\tilde t}^2)^{1/2} \left(m_{\tilde t}^2 - \mu^2\right) + q(m_t^2/m_{\tilde t}^2, \mu^2/m_{\tilde t}^2)^{1/2} \left(m_{\tilde t}^2 - m_t^2 - \mu^2\right)} \to 50\%.
\eeq
where the final approximations for the right-handed stop branching ratios hold in the limit $m_{\tilde t} \gg m_t, \mu$. For the purposes of simplified models, one could either use the full expressions for the right-handed stop branching ratios or the simpler 25\%, 25\%, 50\% approximation, which fails for light stops but is perhaps more in the spirit of simplified models that capture signatures without worrying about the full details of the underlying theory. 

When a decay to a top and Higgsino is forbidden by phase space, these become:
\beq
{\rm If}~m_{\tilde t} < m_t + \mu: & & \nonumber \\
{\rm Br}({\tilde t}_L \to {\tilde H}^+ b_R) & \approx & 100\%. \\
{\rm Br}({\tilde b}_L \to {\tilde H}^0_1 b_R) & \approx & 50\%. \\
{\rm Br}({\tilde b}_L \to {\tilde H}^0_2 b_R) & \approx & 50\%. \\
{\rm Br}({\tilde t}_R \to {\tilde H}^+ b_L) & \approx & 100\%.
\eeq
The two-body decay ${\tilde t}_L \to b_R {\tilde H}^+$ beats the three-body decay ${\tilde t}_L \to W^+ b {\tilde H}^0$ for most values of $\tan \beta$. The three body decay becomes potentially competitive only for $\tan \beta \lsim 2$, as we have checked with BRIDGE~\cite{Meade:2007js}. If small $\tan \beta$ is of interest, e.g. for the NMSSM or $\lambda$SUSY, one should be aware that SDECAY~\cite{Muhlleitner:2003vg} does not compute this three-body decay mode.

\subsection{$SH_u H_d$ Model}
\label{app:shuhd}

The model is 
\beq
W=\lambda S H_u H_d + \frac{m_S}{2}S^2+\frac{\kappa}{3}S^3+\mu H_u H_d.
\eeq
Here $H_u H_d$ denotes the SU(2) symmetric product $H_u^+ H_d^- - H_u^0 H_d^0$. We decouple the bino and wino from our simplified model; the two lightest MSSM neutralinos and chargino are (approximately) Higgsinos. The different Higgsinos are split by dimension-5 operators like $H_u^\dagger T^i \tilde{H}_u H_d^\dagger T^i \tilde{H}_d$ when the bino and wino are integrated out, which we  neglect for the moment. We can decompose the neutral Higgsinos in terms of the two linear combinations
\beq
{\tilde H}^0_\pm = \frac{1}{\sqrt{2}} \left(\tilde{H}^0_u \pm \tilde{H}^0_d\right),
\eeq
in which case the neutral fermion mass terms in the Lagrangian before mixing are three Majorana masses, not all of the same sign:
\beq
-{\cal L}_{\rm mass}  =  m_{\tilde {S}} {\tilde S} {\tilde S} -\frac{\mu}{2} {\tilde H}^0_+ {\tilde H}^0_+ + \frac{\mu}{2} {\tilde H}^0_- {\tilde H}^0_-, 
\eeq
neglecting Higgsino mass shifts of order $m_Z^2/M_{1,2}$ from mixing with the bino and wino. To the first order in $\lambda$,  these states mix after electroweak symmetry breaking into
\beq
\tilde{N}_0 &=& \tilde{S} + \theta_+ \tilde{H}^0_+ + \theta_- \tilde{H}^0_- \nonumber, \\
\tilde{N}_\pm & \approx & {\tilde H}^0_\pm - \theta_\pm \tilde{S}, \nonumber \\
\theta_\pm & \approx & -\frac{\lambda}{2} \frac{v_d \pm v_u}{m_{\tilde {S}} \pm \mu}.
\eeq
If $M_{1,2}$ and $\mu$ all have the same sign, $N_-$ will be the lighter Higgsino eigenstate and $N_+$ the heavier one. In the discussions below, we assume $\mu$ to be positive for simplicity. 

We have computed the partial width to a $Z$-boson final state (checked by comparison to similar calculations in~\cite{Gunion:1987yh,Dreiner:2008tw}) as:
\beq
\Gamma({\tilde N}_\pm \to {\tilde N}_0 + Z) & = & \frac{\theta_\mp^2 q_\pm(m_Z)^{1/2}}{8 \pi v^2 \left|m_\pm\right|} \left(\frac{1}{2} \left(m_\pm^4q_\pm(m_Z) + 3 m_Z^2 (m_\pm^2 + m_0^2 - m_Z^2)\right) + 3 m_\pm m_0 m_Z^2\right) \nonumber \\
& \approx & \frac{\lambda^2 (\cos \beta \mp \sin \beta)^2 q_\pm(m_Z)^{1/2}}{32 \pi (m_{\tilde {S}} \mp \mu)^2 \mu } \left(\frac{1}{2}\left(\mu^4q_\pm(m_Z) + 3 m_Z^2 (\mu^2 + m_{\tilde {S}}^2 - m_Z^2)\right) \mp 3 \mu m_{\tilde {S}} m_Z^2\right), \nonumber \\
\label{eq:neutralinoZ}
\eeq
where $v = 246$ GeV, $m_0, m_+, m_-$ are the masses of $\tilde{N}_0, \tilde{N}_+, \tilde{N}_-$ respectively and 
\beq
q_+(m_Z) &\equiv &q(m_Z^2/m_+^2, m_0^2/m_+^2), \quad q_-(m_Z) \equiv q(m_Z^2/m_-^2, m_0^2/m_-^2).
\eeq
In the second line of Eq.~\ref{eq:neutralinoZ}, we only kept the leading term in an expansion in $\lambda$.
In the limit $\mu \gg m_Z, m_S$ (a limit that is unnatural, but useful for displaying a simpler formula), we can approximate this as
\beq
\Gamma({\tilde N}_\pm \to {\tilde N}_0 + Z) \approx \frac{\lambda^2 \mu}{64 \pi} \left(1 \mp \frac{2 m_{\tilde {S}}}{\mu} \pm \frac{2 m_{\tilde {S}} (m_{\tilde {S}}^2 - m_Z^2)}{\mu^2} + \ldots\right).
\eeq
Since we are most interested in the regime when $m_Z, m_{\tilde {S}},$ and $\mu$ are all near the weak scale, the expression in parentheses will in general be an order-one number.

Two body decays $\tilde{N}_\pm \to \tilde{N}_0 + h$ are mediated by the interaction term in the superpotential in Eq. 1. To the leading order in $\lambda$, the partial widths are given by
\beq
\Gamma({\tilde N}_\pm \to {\tilde N}_0 + h) \approx \frac{\lambda^2 q_\pm(m_h)^{1/2}(\cos \alpha \mp \sin \alpha)^2}{64 \pi \left|m_\pm\right|} \left(m_\pm^2 + 2 m_\pm m_{\tilde {S}} + m_{\tilde {S}}^2 - m_h^2\right).
\eeq

There are also decays to a singlet scalar or pseudoscalar and singlino (previously considered in \cite{Evans:2013jna}), via a mixing of the Higgsino into ${\tilde S}$ which then decays through the superpotential term $\kappa S^3$. These have decay width
\beq
\Gamma({\tilde N}_\pm \to {\tilde N}_0 + s) \approx \frac{\kappa^2 \lambda^2 (v_d \pm v_u)^2 q_{\pm}(m_S)^{1/2}\mu}{32 \pi (m_{\tilde {S}} \pm \mu)^2}  \left(1+\frac{m_{\tilde {S}}^2}{\mu^2}-\frac{m_{S}^2}{\mu^2}+2\frac{m_{\tilde{S}}}{\mu}\right),
\eeq
where we have neglected the splitting between the singlet and singlino and distinguished the singlet mass $m_S$ from the singlino mass $m_{\tilde{S}}$. The decay to pseudoscalar, $\Gamma({\tilde N}_\pm \to {\tilde N}_0 + a)$, has a similar width:
\beq
\Gamma({\tilde N}_\pm \to {\tilde N}_0 + a) \approx \frac{\kappa^2 \lambda^2 (v_d \pm v_u)^2 q_{\pm}(m_S)^{1/2}\mu}{32 \pi (m_{\tilde {S}} \pm \mu)^2}  \left(1+\frac{m_{\tilde {S}}^2}{\mu^2}-\frac{m_{S}^2}{\mu^2}-2\frac{m_{\tilde{S}}}{\mu}\right).
\eeq
For the simplified models in our collider studies, we have not distinguished $a$ from $s$. As a result, we take the approximation of a single real scalar $S$ whose decay width is the sum:
\beq
\Gamma_{\rm simple}({\tilde N}_\pm \to {\tilde N}_0 + S) \approx \frac{\kappa^2 \lambda^2 (v_d \pm v_u)^2 q_{\pm}(m_S)^{1/2}\mu}{16 \pi (m_{\tilde {S}} \pm \mu)^2}  \left(1+\frac{m_{\tilde {S}}^2}{\mu^2}-\frac{m_{S}^2}{\mu^2}\right).
\eeq

The partial width of the two body decay of charged Higgsino is given by 
\beq
\Gamma ( \tilde{C}_1^\pm\to  \tilde{N}_0 + W^\mp) &=& \frac{q(m_W^2/\mu^2, m_0^2/\mu^2)^{1/2}}{8 \pi \mu v^2} \left(\frac{\theta_+^2+\theta_-^2}{2}(\mu^4q(m_W^2/\mu^2, m_0^2/\mu^2)\right. \nonumber \\
&&\left.+3m_W^2(\mu^2+m_0^2-m_W^2))+3(\theta_+^2-\theta_-^2)\mu m_0m_W^2 \right), \\
&\approx&\frac{\lambda^2 q(m_W^2/\mu^2, m_S^2/\mu^2)^{1/2}}{32 \pi \mu \left(\mu^2-m_S^2\right)^2}\left(\left(m_S^2+\mu^2-2 \sin (2\beta) m_S \mu\right)(\mu^4q(m_W^2/\mu^2, m_S^2/\mu^2) \right. \nonumber \\
&&\left.+3m_W^2(\mu^2+m_S^2-m_W^2))+3(-2\mu m_S+ \sin(2\beta)(\mu^2+m_S^2))\mu m_Sm_W^2\right)
\eeq

Three body decay widths scale as
\beq
\Gamma({\tilde H}^\pm \to {\overline f} f' {\tilde H}^0_1) \approx \frac{2}{5 \pi^3} N_c \frac{\delta^5}{v^4} \left(1 - \frac{3}{2} \frac{\delta}{\mu} + \ldots \right).
\eeq
From this we conclude that two-body decays dominate over three-body decays if
\beq
\delta \lsim \left(\frac{5}{128 N_c} \lambda^2 \pi^2 v^4 \mu\right)^{1/5},
\eeq
a condition that is easily fulfilled when $M_{1,2} \gg \mu$.
The three-body decays of $\tilde{C}_1\to W^{\pm *} \tilde{N}_1$ and $\tilde{N}_2 \to \tilde{N}_1 + Z^*$ are calculated in~\cite{Djouadi:2001fa} and is encoded in SDecay.

In Table~\ref{tab:shuhd} we give a sample point and the partial widths of all two body decays.

\begin{table}
\begin{center}
\begin{tabular}{|c|c|}
\hline
\multicolumn{2}{|c|}{$S H_u H_d$} \\
\hline
$m = 80\,\unit{GeV}$ & $m_a = 90\,\unit{GeV}$ \quad $m_s = 103\,\unit{GeV}$  \\
$ \mu = 300\,\unit{GeV}$ &  $m_h = 125\,\unit{GeV}$ \\
$\lambda = -0.02 \quad \kappa = 0.5$ & $\sigma_{s Z} = 0.22 \, \sigma_{h Z}$\\
$\tan \beta = 10 \quad m_A = 700\,\unit{GeV}$ &$ \Gamma_a = 6 \times 10^{-8}\,\unit{GeV}$ \\
$M_1 = 600\,\unit{GeV}$ &  $m_{\tilde s} = 100\,\unit{GeV}$\\
$M_2 = 700\,\unit{GeV}$ & $N_{\tilde s (\tilde H_u, \tilde H_d)} = (0.013, -0.0037)$ \\
$M = -2\,\unit{TeV}$& $N_{\tilde s( \tilde B, \tilde W^0)} = (-0.0011, 0.0017)$\\
\hline 
$\Gamma(\tilde{N}_1 \to \tilde{S} + Z) = 3.3 \times 10^{-4}\,\unit{GeV}$ &$\Gamma(\tilde{N}_1 \to \tilde{S} + h) = 5.1 \times 10^{-4}\,\unit{GeV}$ \\
$\Gamma(\tilde{N}_1 \to \tilde{S} + S) = 5.7 \times 10^{-4}\,\unit{GeV}$ & $\Gamma(\tilde{C}_1 \to \tilde{S} + W) = 9.6 \times 10^{-4}\,\unit{GeV}$\\
$\Gamma(\tilde{N}_2 \to \tilde{S} + Z) = 5.7 \times 10^{-4}\,\unit{GeV}$ &$\Gamma(\tilde{N}_2 \to \tilde{S} + h) = 1.3 \times 10^{-4}\,\unit{GeV}$ \\
$\Gamma(\tilde{N}_2 \to \tilde{S} + S) = 2.1 \times 10^{-4}\,\unit{GeV}$ & \\
\hline
\end{tabular}
\end{center}
\caption{A benchmark point for the $S H_u H_d$ model. To lift the Higgs mass above the experimental limit (even if stops are light), we add $\left(H_u H_d\right)^2/M$ to the superpotential~\cite{Dine:2007xi}. We also list the partial widths of all two-body decays. }\label{tab:shuhd}
\end{table}

\subsection{$SY {\overline Y}$ Model}
\label{app:syy}
The model is 
\beq
W=\lambda S Y {\overline Y}+ \frac{m_S}{2}S^2+m_YY {\overline Y},
\eeq
where $Y$ and ${\overline Y}$ transform as $5$ and ${\overline 5}$ under SM gauge group. In this model, the couplings between SM singlet $S$ and MSSM particles are generated through loop of $Y$ and ${\overline Y}$. Specifically, at one-loop level, integrating out $Y$ and ${\overline Y}$ generates the operator 
\beq
c \int d^2 \theta S W^\alpha W_\alpha + {\rm h.c.}, 
\eeq
where $W^\alpha$ is the field strength. In terms of component fields, this leads to operators such as $\tilde{S} \sigma^{\mu\nu} G^{\mu\nu a} \tilde{g}^a$ and $s G^2$. The coefficient $c$ has been worked out via low energy theorem in~\cite{Fan:2012jf}
\beq
c=\sum_i \lambda \frac{\sqrt{2} \alpha b_i}{16\pi m_Y},
\label{eq:coe}
\eeq
where the sum runs over the contributions to the gauge coupling beta function coefficients from all heavy fields running in the loop. 

In the simplified models, stops decay through an off-shell gluino to $t g \tilde{S}$ or through an off-shell electroweakino to $t \gamma (Z) \tilde{S}$ or $b W \tilde{S}$. If the gauginos are similar in masses and kinematically allowed, the QCD decay channel $t g \tilde{S}$ will be the dominate one. The two neutral Higgsinos will decay through mixing with neutral wino or bino to $Z \tilde{S}$ or $\gamma \tilde{S}$ with the ratio of the partial widths
\beq
\frac{\Gamma(\tilde{N}_i \to Z + \tilde{S})}{\Gamma(\tilde{N}_i \to \gamma + \tilde{S})} \propto \left( \frac{N_{i2}c_W c_2 - N_{i1}s_W c_1}{N_{i2}s_W c_2 + N_{i1}c_W c_1} \right)^2,
\eeq
where $N_{ij}$ is the matrix that diagnalize the MSSM neutralino mass matrix, $c_W (s_W)$ are the cosine (sine) of the Weinberg angle and $c_2 (c_1)$ is the coefficient in Eq.~\ref{eq:coe} for $SU(2)_W$ ($U(1)_Y$). 
One could see that the branching ratios depends on two linearly independent combinations of $N_{i1}$ and $N_{i2}$, which are determined by free parameters, the bino and wino masses. Thus we'll simply treat the branching ratios as a free parameter. 
The charged Higgsino will decay to $W \tilde{S}$ again through mixing with charged winos. There are still the usual 3-body transitions within the Higgsino multiplet, but these scale as $1/M_{1,2}^5$ at small splittings while the loop processes we consider here scale as $1/(m_Y^2 M_{1,2}^2)$ and can dominate when the (electroweak) $Y$ states are relatively light and the bino and wino are above a TeV. As in the previous model, $\tilde{S} \to s + \tilde{a} (\tilde{G})$. Finally, the scalar $s$ decays back to the SM, $s \to gg, \gamma\gamma$. As a benchmark case, we focus on the scenario where the Higgsino mixes dominantly with the wino, so that ${\rm Br}({\tilde H}^0 \to Z + {\tilde S}) = c_W^2$ and ${\rm Br}({\tilde H}^0 \to \gamma + {\tilde S}) = s_W^2$.

\section{Validating Simulated LHC Searches}
\label{app:validation}

\subsection{ATLAS multijet study}

In this subsection we validate a simulation of the ATLAS multijet study~\cite{TheATLAScollaboration:2013xia,Aad:2015lea}, focusing on the jet-counting studies. We simulate the same signals tabulated in the ATLAS note and compare our expected number of events to their results in Tables~\ref{tab:compare0tag} and~\ref{tab:comparetag}.

\rowcolors{2}{blue!15}{white}
\begin{table}[H]
\begin{center}
\begin{tabular}{llllll}
Channel & Table & Model & ATLAS number & Our number & Ratio\\
\hline
7j80, 0b & 6 & ${\tilde g}$ (400), $\chi$ (300), 10q, $\lambda''_{112}$ & 9000 $\pm$ 4000 & 17900 & 2.0 \\
\hline
7j100, 0b & 6 & ${\tilde g}$ (400), $\chi$ (50), 10q, $\lambda''_{112}$ & 1400 $\pm$ 800 & 3300 & 2.4 \\
7j100, 0b & 6 & ${\tilde g}$ (600), $\chi$ (300), 10q, $\lambda''_{112}$ & 1700 $\pm$ 900 & 2600 & 1.5 \\
\hline
7j120, 0b & 2 & ${\tilde g}$ (500), 6q, $\lambda''_{112}$ & 600 $\pm$ 230 & 740 & 1.2 \\
7j120, 0b & 2 & ${\tilde g}$ (600), 6q, $\lambda''_{112}$ & 410 $\pm$ 100 & 550 & 1.3 \\
7j120, 0b & 6 & ${\tilde g}$ (800), $\chi$ (300), 10q, $\lambda''_{112}$ & 380 $\pm$ 90 & 330 & 0.9\\
\hline
7j140, 0b & 6 & ${\tilde g}$ (1000), $\chi$ (300), 10q, $\lambda''_{112}$ & 50 $\pm$ 13 & 45 & 0.9\\
7j140, 0b & 6 & ${\tilde g}$ (1200), $\chi$ (600), 10q, $\lambda''_{112}$ & 28 $\pm$ 4 & 27 & 1.0\\
\hline
6j180, 0b & 6 & ${\tilde g}$ (1000), $\chi$ (50), 10q, $\lambda''_{112}$ & 40 $\pm$ 6 & 31 & 0.8 \\
\hline
7j180, 0b & 2 & ${\tilde g}$ (800), 6q, $\lambda''_{112}$ & 13 $\pm$ 4 & 13 & 1.0 \\
7j180, 0b & 2 & ${\tilde g}$ (1000), 6q, $\lambda''_{112}$ & 6.8 $\pm$ 2.3 & 6.2 & 0.9 \\
7j180, 0b & 2 & ${\tilde g}$ (1200), 6q, $\lambda''_{112}$ & 2.7 $\pm$ 0.5 & 2.3 & 0.9 \\
7j180, 0b & 6 & ${\tilde g}$ (1000), $\chi$ (600), 10q, $\lambda''_{112}$ & 10 $\pm$ 5 & 11 & 1.1\\
7j180, 0b & 6 & ${\tilde g}$ (1200), $\chi$ (50), 10q, $\lambda''_{112}$ & 1.9 $\pm$ 1.0 & 2.0 & 1.1\\
7j180, 0b & 6 & ${\tilde g}$ (1200), $\chi$ (300), 10q, $\lambda''_{112}$ & 3.2 $\pm$ 1.4 & 4.2 & 1.3\\
\hline
\end{tabular}
\end{center}
\caption{Validation: cases with 0 $b$-tags}\label{tab:compare0tag}
\end{table}

\rowcolors{2}{blue!15}{white}
\begin{table}[H]
\begin{center}
\begin{tabular}{llllll}
Channel & Table & Model & ATLAS number & Our number & Ratio\\
\hline
7j80, 1b & 4 & ${\tilde g}$ (500), 6q, $\lambda''_{312}$ & 4600 $\pm$ 800 & 6750 & 1.5\\
\hline
7j100, 1b & 4 & ${\tilde g}$ (600), 6q, $\lambda''_{312}$ & 940 $\pm$ 190 & 1027 & 1.1 \\
7j100, 1b & 6 & ${\tilde g}$ (600), $\chi$ (50), 10q, $\lambda''_{112}$ & 510 $\pm$ 140 & 650 & 1.3 \\
\hline
7j120, 1b & 3 & ${\tilde g}$ (600), 6q, $\lambda''_{113,123}$ & 300 $\pm$ 60 & 415 & 1.4 \\
7j120, 1b & 3 & ${\tilde g}$ (800), 6q, $\lambda''_{113,123}$ & 131 $\pm$ 25 & 143 & 1.1 \\
7j120, 1b & 4 & ${\tilde g}$ (800), 6q, $\lambda''_{312}$ & 108 $\pm$ 18 & 145 & 1.3 \\
7j120, 1b & 4 & ${\tilde g}$ (1000), 6q, $\lambda''_{312}$ & 42 $\pm$ 6 & 48 & 1.1 \\
7j120, 1b & 6 & ${\tilde g}$ (800), $\chi$ (50), 10q, $\lambda''_{112}$ & 107 $\pm$ 31 & 93 & 0.9 \\
\hline
7j180, 1b & 3 & ${\tilde g}$ (1000), 6q, $\lambda''_{113,123}$ & 4.4 $\pm$ 1.0 & 4.6 & 1.0 \\
7j180, 1b & 3 & ${\tilde g}$ (1200), 6q, $\lambda''_{113,123}$ & 1.86 $\pm$ 0.31 & 1.8 & 1.0 \\
7j180, 1b & 4 & ${\tilde g}$ (1200), 6q, $\lambda''_{312}$ & 1.3 $\pm$ 0.4 & 1.5 & 1.2\\
\hline
\end{tabular}
\end{center}
\caption{Validation: cases with 1 $b$-tag}\label{tab:comparetag}
\end{table}

\rowcolors{2}{blue!15}{white}
\begin{table}[H]
\begin{center}
\begin{tabular}{llllll}
Channel & Table & Model & ATLAS number & Our number & Ratio\\
\hline
7j80, 2b & 3 & ${\tilde g}$ (500), 6q, $\lambda''_{113,123}$ & 1900 $\pm$ 400 & 3050 & 1.6 \\
7j80, 2b & 5 & ${\tilde g}$ (500), 6q, $\lambda''_{313,323}$ & 3600 $\pm$ 600 & 6100 & 1.7 \\
7j80, 2b & 5 & ${\tilde g}$ (600), 6q, $\lambda''_{313,323}$ & 2300 $\pm$ 400 & 3200 & 1.4  \\
\hline
7j120, 2b & 5 & ${\tilde g}$ (800), 6q, $\lambda''_{313,323}$ & 94 $\pm$ 15 & 126 & 1.3 \\
7j120, 2b & 5 & ${\tilde g}$ (1000), 6q, $\lambda''_{313,323}$ & 37 $\pm$ 6 & 41 & 1.1 \\
\hline
7j140, 2b & 5 & ${\tilde g}$ (1200), 6q, $\lambda''_{313,323}$ & 5.5 $\pm$ 1.0 & 6.3 & 1.1\\
\hline
\end{tabular}
\end{center}
\caption{Validation: cases with 2 $b$-tags}\label{tab:prior}
\end{table}

\subsection{ATLAS Stop Search with $Z + b + \met$}

\begin{figure}[H]\begin{center}
\includegraphics[width=0.5\textwidth]{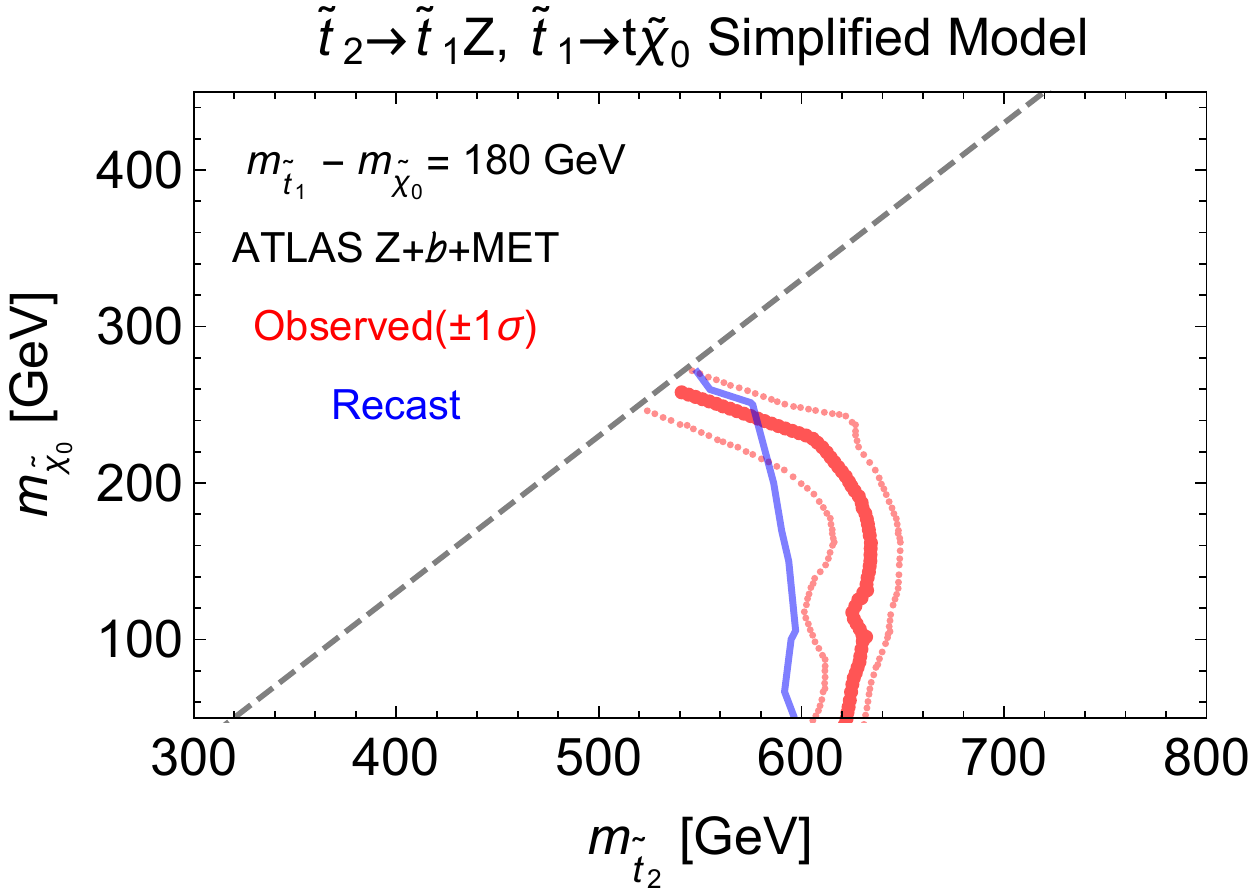}
\end{center}
\caption{Validation of our simulation of the ATLAS bound on $pp \to {\tilde t}_2 {\tilde t}^*_2,~{\tilde t}_2 \to Z {\tilde t}_1$, ${\tilde t}_1 \to t {\tilde \chi}^0_1$ from Figure~4 of ref.~\cite{Aad:2014mha}. It is assumed that $m_{{\tilde t}_1} - m_{{\tilde \chi}_1} = 180$ GeV. The gray diagonal line is the kinematic constraint $m_{{\tilde t}_2} - m_{{\tilde t}_1} = m_Z$. The red curves are the ATLAS constraints (central value and 1$\sigma$ deviation) and the blue line is our estimated exclusion with our estimated efficiency.}
\label{fig:validateatlaszbmet}
\end{figure}

Here we validate the results of our simulation of ref.~\cite{Aad:2014mha}. We compare our estimates to the ATLAS results in Figure~\ref{fig:validateatlaszbmet}.

\subsection{CMS $t\bar{t}$ + jets}
\label{app:ttbarplusjets}

\begin{figure}[!h]\begin{center}
\includegraphics[width=0.45\textwidth]{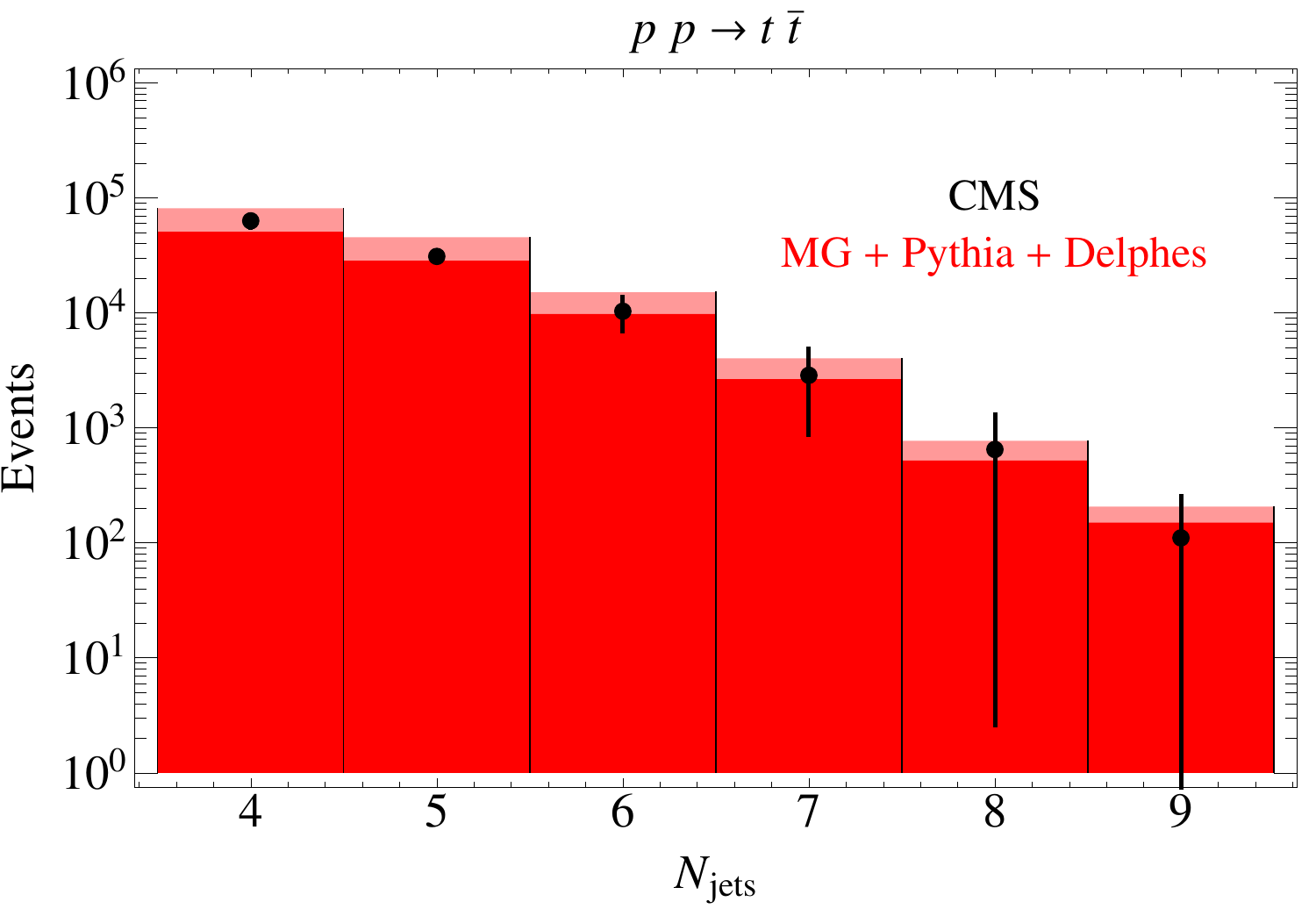} \quad \includegraphics[width=0.45\textwidth]{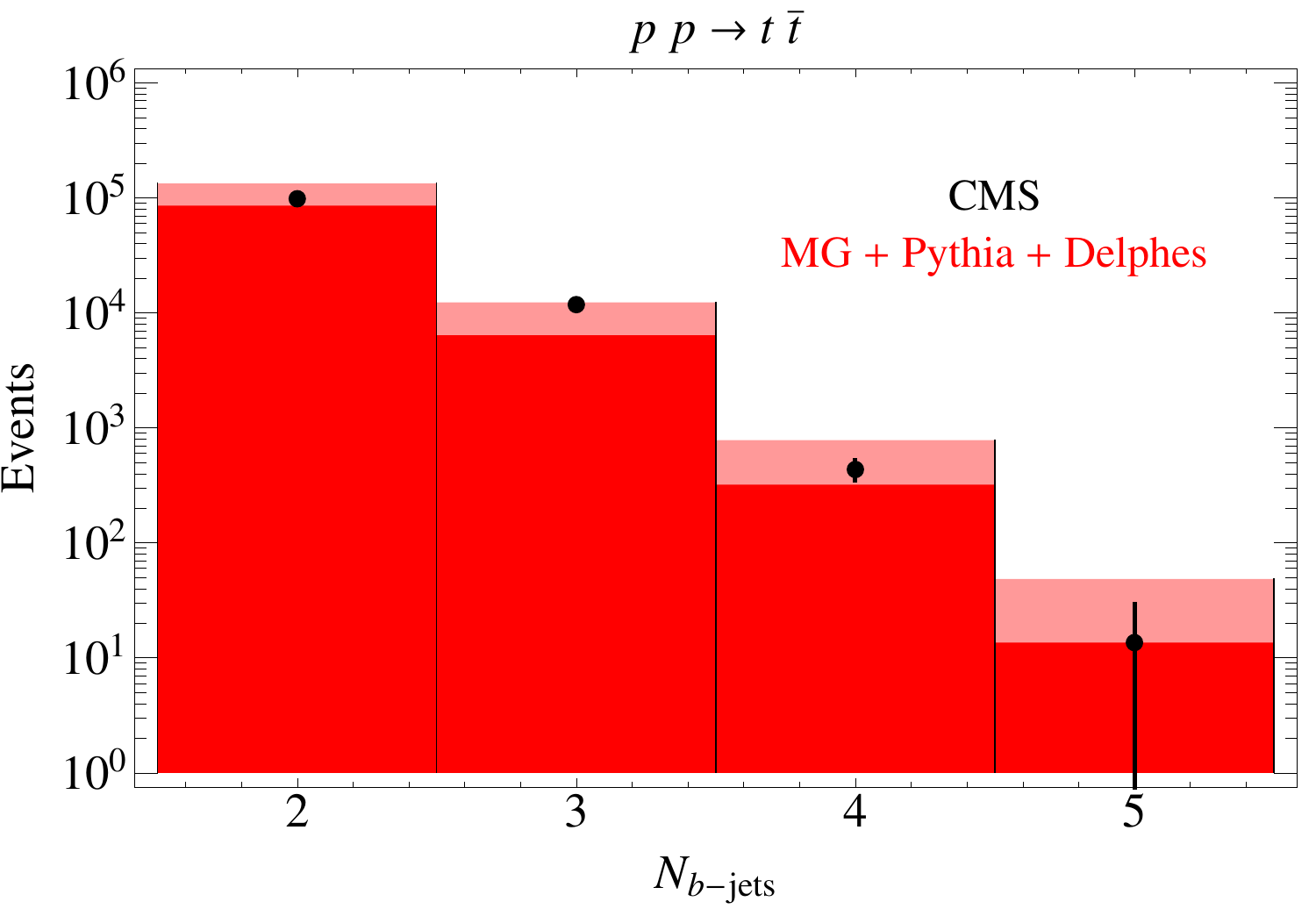}
\end{center}
\caption{Validation of our simulation of the CMS jet multiplicity distributions in semi-leptonic top decays from~\cite{Khachatryan:2015oqa}, both with (right) and without (left) $b$-tags. Light shaded regions reflect an estimate of our systematic uncertainty in modeling CMS detector efficiencies.}
\label{fig:validatettbarjets}
\end{figure}

We have validated a simulation of the jet multiplicity distributions in semi-leptonic top decays from the CMS study~\cite{Khachatryan:2015oqa}, both with and without $b$-tags.  We compare our simulation to the CMS results in Figure~\ref{fig:validatettbarjets}.  At large $b$-jet multiplicities, these distributions depend sensitively on the per-$b$-jet tagging efficiency, $\epsilon_b$.  As an estimate of the systematic uncertainty in our detector simulation, we take a $p_T$- and $\eta$-independent value for $\epsilon_b$ and vary it from $60\%$ (the fiducial value used in the study) down to $50\%$.  The resulting uncertainty on the jet multiplicity distribution is shown shaded in lighter red in the figure.  Note that, as these distributions are defined after cuts which include $b$-tag requirements (to ensure a top-rich sample), the un-tagged jet distribution is also sensitive to $\epsilon_b$.

\subsection{CMS Stop Search with $h \rightarrow \gamma \gamma$}

We have validated a simulation of the CMS study~\cite{Chatrchyan:2013mya}.  Our results for the three benchmark stop / neutralino mass points they considered are shown in Table~\ref{tab:comparehgamgam}.

\rowcolors{2}{blue!15}{white}
\begin{table}[!h]
\begin{center}
\begin{tabular}{llllllll}
$(m_{{\tilde t}_R}, m_{{\tilde \chi}^0_1})$ & (i) $\geq 3 b$ & & (ii) $m_{b{\overline b}} \approx m_h$ & & (iii) $m_{b{\overline b}}$ other  \\
& CMS & us & CMS & us & CMS & us \\
\hline
$(350, 135)$ & 10.7 & 8.4 & 2.0 & 0.6 & 6.8 & 2.6\\
$(300, 290)$ & 2.1 & 2.9 & 10.1 & 9.8 & 3.9 & 4.1 \\
$(400, 300)$ & 4.0 & 3.8 & 1.4 & 0.7 & 2.8 & 1.4\\
\hline
\end{tabular}
\end{center}
\caption{Validation: CMS search for ${\tilde t}_R \to b {\tilde \chi}^+_1,t{\tilde \chi}^0_1$ followed by NLSP decay ${\tilde \chi}^0_1 \to h {\tilde G}$ with $h \to \gamma\gamma$~\cite{Chatrchyan:2013mya}.}\label{tab:comparehgamgam}
\end{table}

Our results are generally a good approximation to the CMS results, especially for the points with higher signal efficiencies. In a few bins we fall short of the CMS estimate by more than a factor of two. Based on these results, we expect that our estimated exclusion will be somewhat weaker than the true CMS exclusion.

\subsection{CMS Same-Sign Dileptons and Jets}

\begin{figure}[!h]\begin{center}
\includegraphics[width=0.5\textwidth]{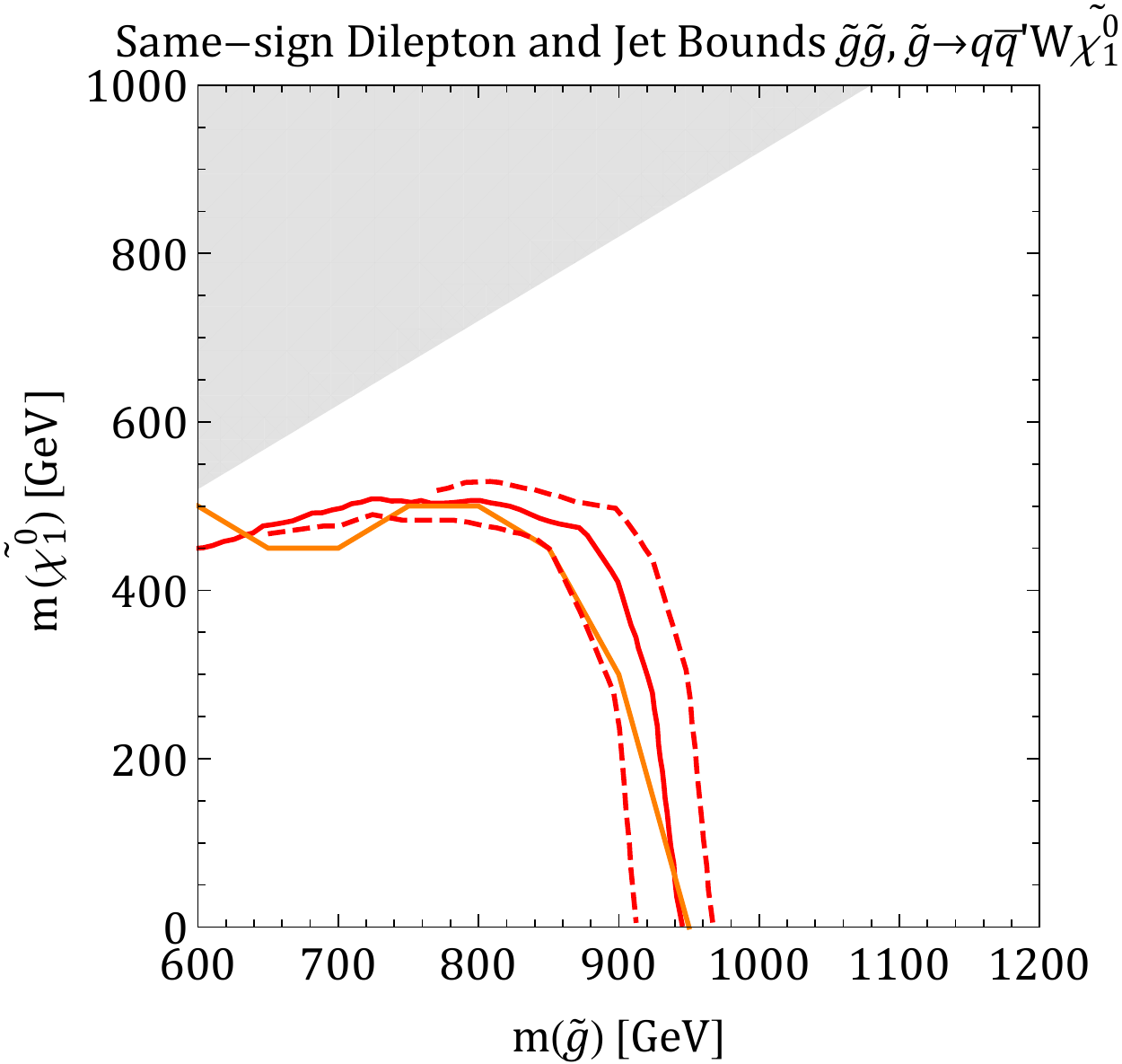}
\end{center}
\caption{Validation of our simulation of the CMS bound on $pp \to {\tilde g}{\tilde g},~{\tilde g} \to q {\overline q}' W {\tilde \chi}^0_1$ from the right panel of Figure~7 of ref.~\cite{Chatrchyan:2013fea}. The red points are digitized from the CMS bound and the orange curve is our simulated result.}
\label{fig:validatecmssus13013}
\end{figure}

We have validated the CMS search for events with same-sign dileptons and jets~\cite{Chatrchyan:2013fea}. This involves several search regions with varying jet multiplicities, $H_T$ requirements, and $\met$ constraints. Some of the bins involve relatively low $\met$ (50 to 120 GeV) which might be fulfilled by neutrinos in cascade decays in Stealth SUSY models.  We show a comparison of our estimates with the CMS results in Figure~\ref{fig:validatecmssus13013}.

\subsection{ATLAS Many Jets and $\met$}

\begin{figure}[!h]\begin{center}
\includegraphics[width=0.5\textwidth]{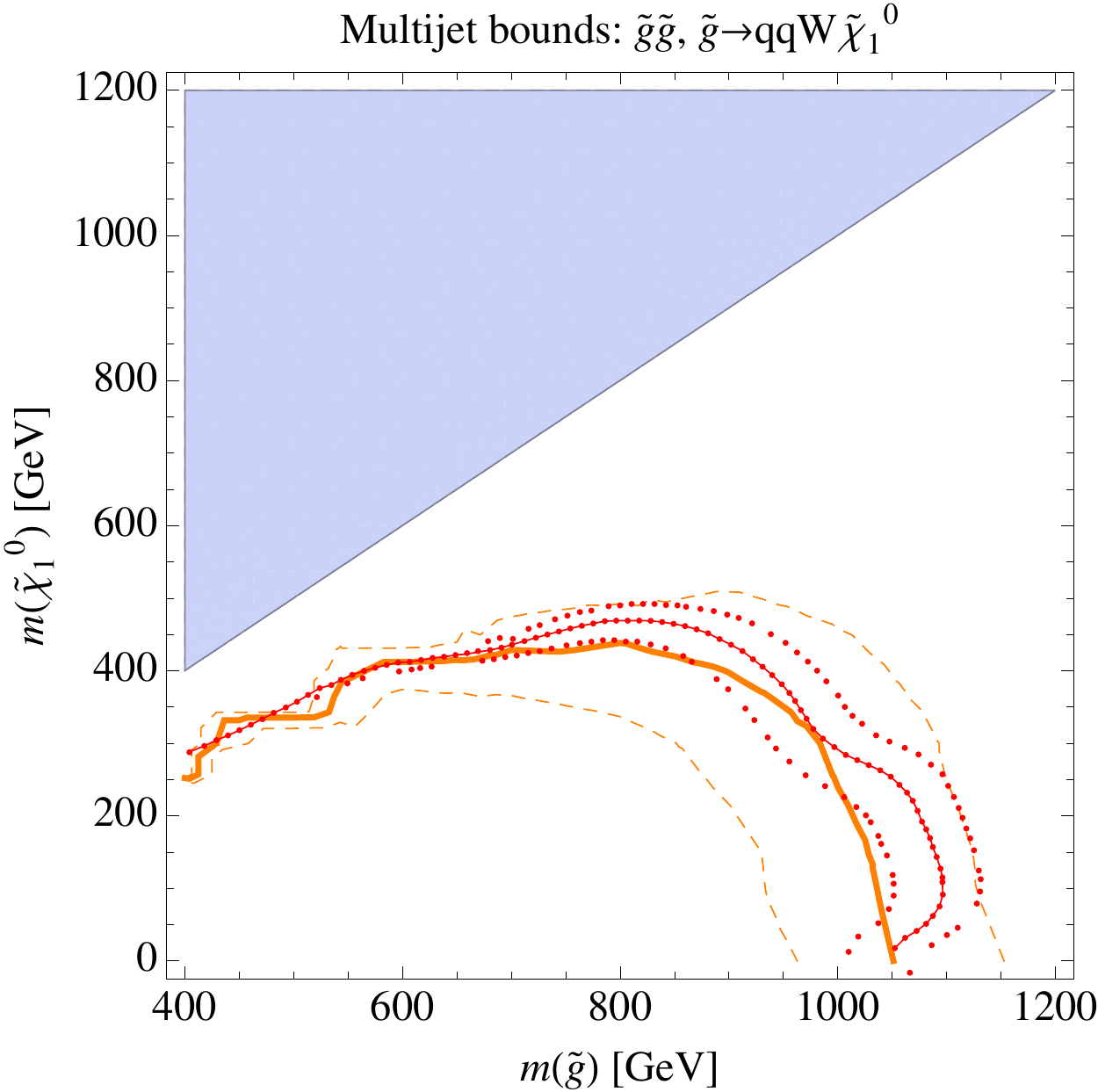}
\end{center}
\caption{Validation of our simulation of the ATLAS bound on events with many jets and $\met$. This is for the simplified model of gluino pair production with ${\tilde g} \to q {\overline q}' W {\tilde \chi}^0_1$, with a chargino mass fixed at $m_{\tilde \chi^\pm} = (m_{\tilde g} + m_{{\tilde \chi}^0_1})/2$, as shown in the left panel of Figure~11 of ref.~\cite{Aad:2013wta}. The red dots plot the ATLAS bound, while the thick orange curve is our estimated exclusion and the dashed orange curves vary our estimated efficiency by a factor of 2 in both directions.}
\label{fig:validateatlas054}
\end{figure}

We have validated a simulation of the ATLAS search for large jet multiplicity events with missing transverse momentum~\cite{Aad:2013wta}. We show a comparison of our estimates with the ATLAS results in Figure~\ref{fig:validateatlas054}.

\subsection{CMS Multilepton Searches}
\begin{figure}[!h]\begin{center}
\includegraphics[width=0.5\textwidth]{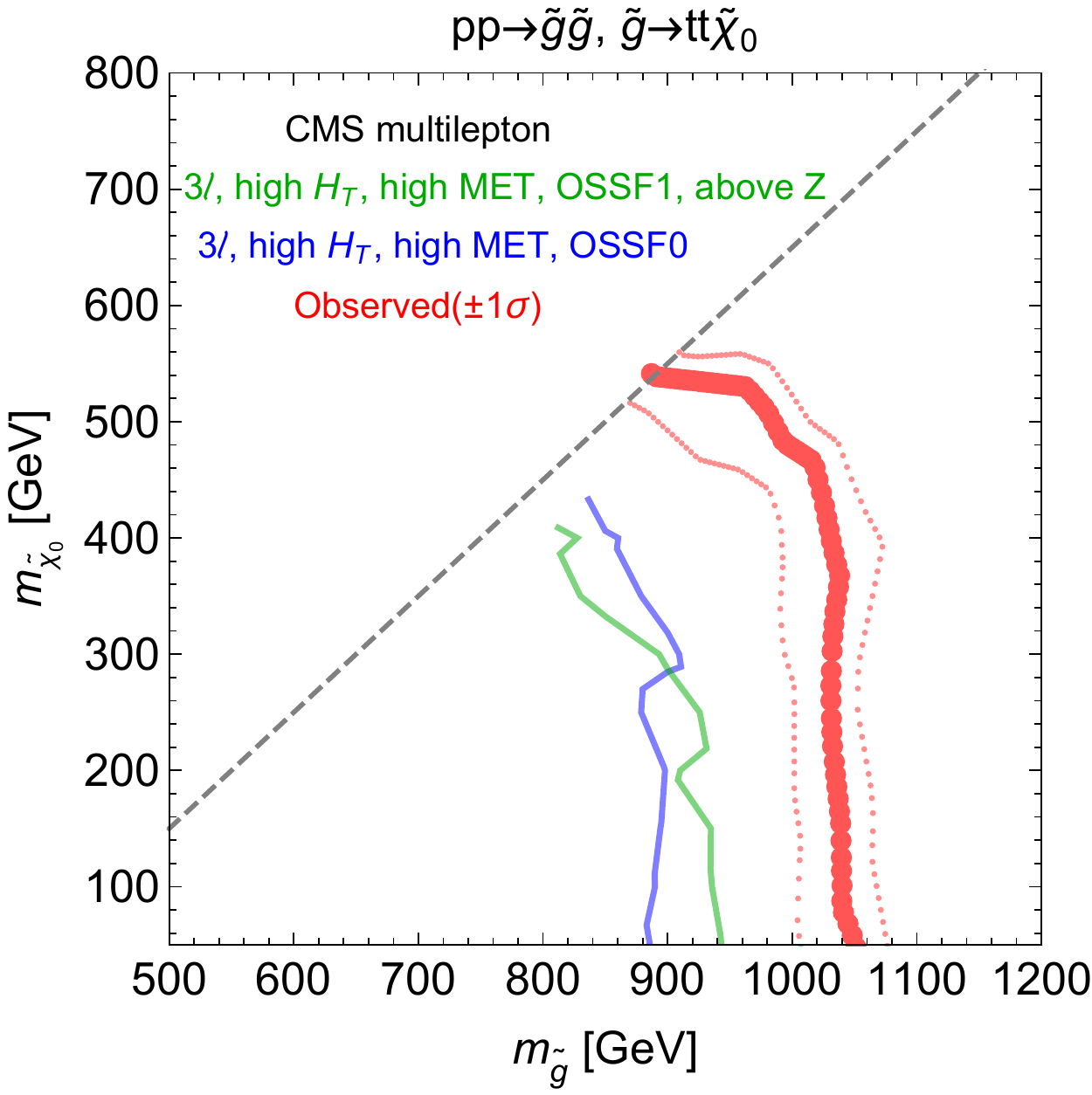}
\end{center}
\caption{Validation of our simulation of the CMS bound on events with many leptons and $\met$ \cite{Chatrchyan:2014aea}. This is for the simplified model of gluino pair production with ${\tilde g} \to t {\overline t} \chi^0_1$. The red dots plot the CMS bound, while the thick blue and green curves are our estimated exclusion from an individual search region. }
\label{fig:validatecmsmultilepton}
\end{figure}

We have validated a simulation of the CMS search for multilepton final states with three or more leptons~\cite{Chatrchyan:2014aea}. We show a comparison of our estimates with the CMS results in Figure~\ref{fig:validatecmsmultilepton}.

\subsection{ATLAS Same-Sign Dileptons and Trileptons}
\begin{figure}[!h]\begin{center}
\includegraphics[width=0.9\textwidth]{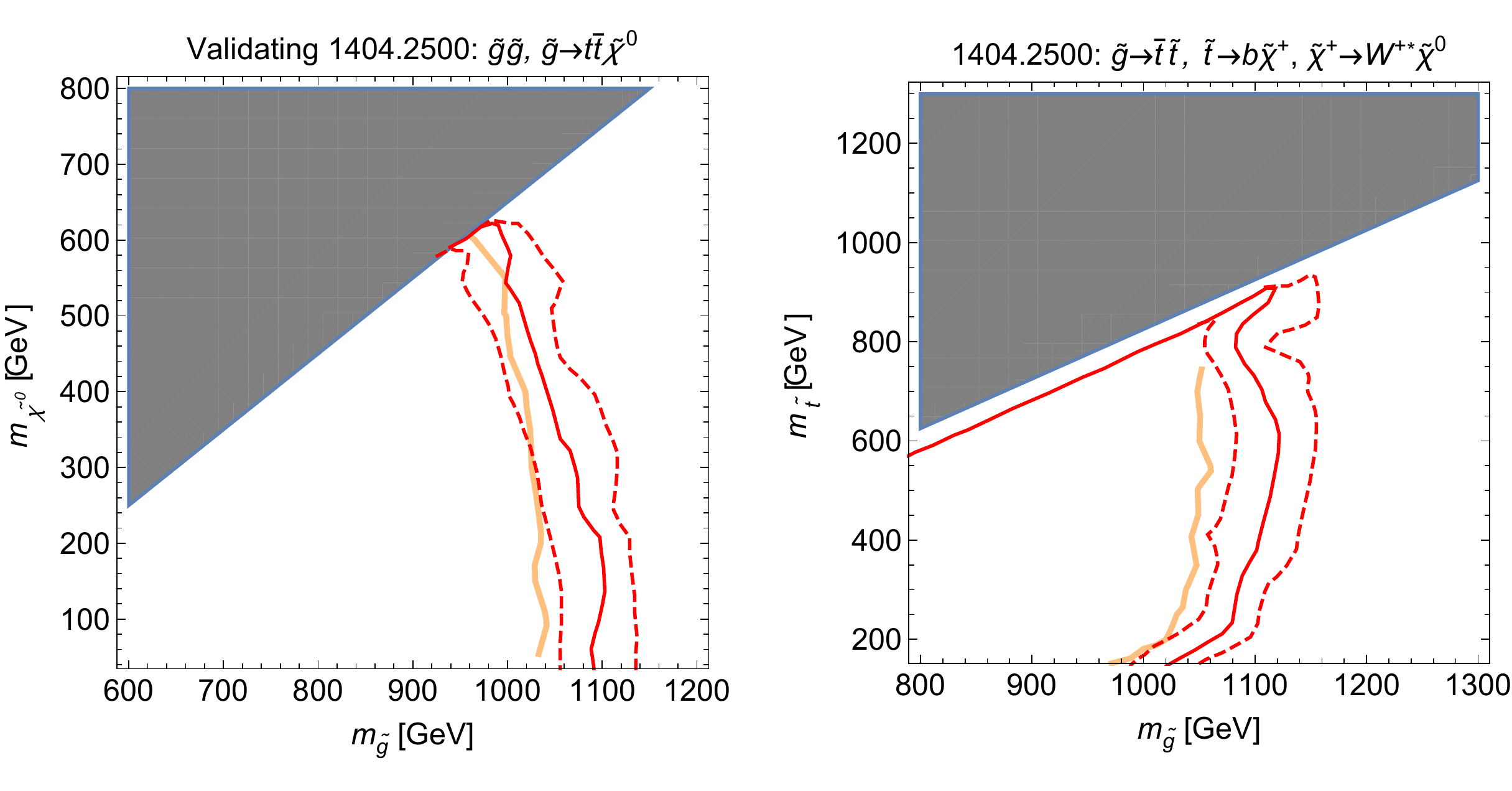}
\end{center}
\caption{Validation of our simulation of the ATLAS bound on events with jets and at least two same-sign leptons or trileptons \cite{Aad:2014pda}. Left: the simplified model of gluino pair production with ${\tilde g} \to t {\overline t}  {\tilde \chi}^0_1$. Right: gluino pair production with ${\tilde g} \to {\overline t} {\tilde t}, {\tilde t} \to b {\tilde \chi}^+, {\tilde \chi}^+ \to W^{+*} {\tilde \chi}^0$ (at fixed masses $m_{{\tilde \chi}^0} = 60~{\rm GeV}$ and $m_{{\tilde \chi}^+} = 118~{\rm GeV}$). The red dots plot the ATLAS bound, while the thick orange curve is our estimated exclusion using CheckMATE.}
\label{fig:validatecheckmate}
\end{figure}

An analysis code for an ATLAS search for events with jets and at least two same-sign leptons or trileptons \cite{Aad:2014pda} is contained in the CheckMATE package without validation. We have validated the code and show a comparison of our estimates with the ATLAS results in Figure~\ref{fig:validatecheckmate}.

\newpage

\bibliography{ref}
\bibliographystyle{utphys}
\end{document}